\documentclass[aps,pra,twocolumn,amssymb,amsfonts,floatfix,superscriptaddress,longbibliography,notitlepage,footinbib]{revtex4-1}
\usepackage[utf8]{inputenc}
\pdfoutput=1
\usepackage{color}
\usepackage{float}
\usepackage{xcolor}
\usepackage{mathtools,amsmath,amsthm,amsfonts,amssymb}
\usepackage{commath}
\usepackage{physics}
\usepackage{bm}
\usepackage[pdftex]{graphicx}
\usepackage[colorlinks=false]{hyperref}
\usepackage{enumitem}
\usepackage{soul}

\begin{document}

\title{Breakdown of the Quantum Distinction of Regular and Chaotic Classical Dynamics in Dissipative Systems}

\author{David Villase\~nor}
\affiliation{Instituto de Investigaciones en Matem\'aticas Aplicadas y en Sistemas, Universidad Nacional Aut\'onoma de M\'exico, C.P. 04510 Mexico City, Mexico}
\author{Lea F. Santos} 
\affiliation{Department of Physics, University of Connecticut, Storrs, Connecticut 06269, USA}
\author{Pablo Barberis-Blostein}
\affiliation{Instituto de Investigaciones en Matem\'aticas Aplicadas y en Sistemas, Universidad Nacional Aut\'onoma de M\'exico, C.P. 04510 Mexico City, Mexico}

\begin{abstract}
Quantum chaos has recently received increasing attention due to its relationship with experimental and theoretical studies of nonequilibrium quantum dynamics, thermalization, and the scrambling of quantum information. In an isolated system, quantum chaos refers to properties of the spectrum that emerge when the classical counterpart of the system is chaotic. However, despite experimental progress leading to longer coherence times, interactions with an environment can never be neglected, which calls for a definition of quantum chaos in dissipative systems. Advances in this direction were brought by the Grobe-Haake-Sommers (GHS) conjecture, which connects chaos in a dissipative classical system with cubic repulsion of the eigenvalues of the quantum counterpart and regularity with linear level repulsion. Here, we show that the GHS conjecture does not hold for the open Dicke model, which is a spin-boson model of experimental interest. We show that the onset of cubic level repulsion in the open quantum model is not always related with chaotic structures in the classical limit. This result challenges the universality of the GHS conjecture and raises the question of what is the source of spectral correlations in open quantum systems.
\end{abstract}

\maketitle

Chaos in isolated classical systems entails extreme sensitivity to
initial conditions and mixing~\cite{OttBook,Gutzwiller1990book}. The subject is extremely broad, with various applications in science and engineering, but a missing element of the theory is the implications of classical chaos to quantum systems. The most accepted definition of quantum chaos is based on a conjecture, the Bohigas-Giannoni-Schmit conjecture~\cite{Bohigas1984,Casati1980}, that states that the energy level statistics of isolated quantum systems whose classical counterparts are chaotic agrees with universal properties of random matrices, that is, the eigenvalues are correlated~\cite{MehtaBook,HaakeBook,WimbergerBook,Guhr1998} and the eigenstates are similar to random vectors~\cite{Berry1977JPA,Borgonovi2006}. 
In contrast, the eigenvalues of quantum systems with regular classical counterparts are uncorrelated, which became known as the Berry-Tabor conjecture~\cite{Berry1977}. Quantum chaos thus refers to spectral signatures  of chaotic behavior in the classical limit.

The Bohigas-Giannoni-Schmit conjecture is supported by several numerical studies of one-body and few-body quantum systems with a chaotic classical counterpart~\cite{Bohigas1984,Seligman1984,Seligman1985,Friedrich1989,Haake1987,Chaudhury2009,Izrailev1990,Santhanam2022}, and the recent analysis in Ref.~\cite{Wang2022} suggests that it could be extended to classical systems that are ergodic, but not necessarily chaotic. The widespread numerical evidence of the validity of the Bohigas-Giannoni-Schmit conjecture has encouraged extending the use of level statistics as a probe of quantum chaos to many-body quantum systems, including cases that do not have a well-defined classical limit, such as interacting many-body spin-1/2 models~\cite{Hsu1993,Santos2020}. Exceptions to the Berry-Tabor conjecture have been found, but they are usually not robust~\cite{Casati1985,Relano2004} or are associated with systems that are not physical~\cite{Benet2003}.

Given the overall success of the Bohigas-Giannoni-Schmit and the  Berry-Tabor conjectures as tools to describe chaos and regularity in isolated quantum systems, and the fact that isolated systems are idealizations, a natural question is how these conjectures can be extended to open systems. This question has recently received a great deal of attention due to the relationship between quantum chaos and several topics of theoretical and experimental interest, including thermalization~\cite{Borgonovi2016,Dalessio2016}, scrambling of quantum information~\cite{Landsman2019}, difficulties to control and localize many-body quantum systems~\cite{Suntajs2020}, and scattering near the boundaries of black holes~\cite{Jensen2016}. The current employed notion of quantum chaos in dissipative systems~\cite{Denisov2019,Akemann2019,Sa2020,Hamazaki2020,Li2021,Rubio2022,Garcia2022,Kawabata2023,Sa2023,Garcia2023,Ferrari2023Arxiv}  follows the findings from Ref.~\cite{Grobe1988}, where the quantum and classical analysis of a periodically kicked top with damping suggested a relationship between chaos in open classical systems, characterized by the appearance of chaotic attractors~\cite{OttBook,Guckenheimer1983book,Shivamoggi2014book,Smyrlis1991,Papageorgiou1991}, and level repulsion as in the Ginibre ensembles of non-Hermitian random matrices for the quantum counterparts. It was found in Ref.~\cite{Grobe1988} that the presence of a chaotic attractor in the classical limit results in cubic level repulsion for the complex spectrum, while the repulsion becomes linear if the attractor is simple. The extension of this correspondence to other dissipative systems constitutes what became known as the Grobe-Haake-Sommers (GHS) conjecture~\cite{Grobe1988,Grobe1989,Akemann2019}. The conjecture is employed to the complex eigenvalues of the Liouvillian of Markovian Lindblad master equations~\cite{BreuerBook,CarmichaelBook1993,CarmichaelBook2002} and also to effective non-Hermitian Hamiltonians used to treat
dissipation~\cite{Jaiswal2019,Hamazaki2019,Sa2020}. Similar to the
case of isolated systems, the GHS conjecture has also been used to
describe dissipative many-body quantum systems without well-defined
classical counterparts~\cite{Akemann2019,Hamazaki2020,Sa2020,Rubio2022,Garcia2022,Kawabata2023,Sa2023,Garcia2023}.

Here, we investigate the validity of the GHS conjecture for the open Dicke model, where dissipation is caused by photon leakage. The model describes the interaction between a set of particles and a single-mode electromagnetic field in a cavity, and has been experimentally realized~\cite{Jaako2016,Baden2014,Klinder2015PNAS,Zhiqiang2017,Zhang2018,Cohn2018,Safavi2018}. It was first introduced to study superradiance~\cite{Dicke1954,Garraway2011,Kirton2019} and has since been applied to various different subjects~\cite{Villasenor2024ARXIV}, such as classical and quantum chaos, excited state quantum phase transitions, and quantum batteries. Level repulsion as in Ginibre ensembles was verified for the quantum open isotropic Dicke model in Ref.~\cite{Villasenor2024}, while chaotic attractors were found for the classical dissipative anisotropic model in Ref.~\cite{Stitely2020}. 

The present work provides a systematic comparison between classical chaos and level statistics in the open Dicke model. We show that there is a wide range of system parameters for which cubic level repulsion is observed, while the dissipative classical dynamics does not reveal chaotic attractors, which contradicts the GHS conjecture. This finding suggests that the GHS conjecture is not universal and the characterization of quantum chaos in dissipative systems needs to be reconsidered.

{\it Quantum Dicke model---}
The Dicke model describes $\mathcal{N}$ two-level atoms with energy splitting $\omega_{0}$ interacting with a single-mode electromagnetic field of frequency $\omega$. The Dicke Hamiltonian for $\hbar=1$ is
\begin{align}
    \label{eq:DickeHamiltonian}
    \hat{H}_{\text{D}} = & \; \omega\hat{a}^{\dagger}\hat{a} + \omega_{0}\hat{J}_{z} + \frac{\gamma_{-}}{\sqrt{\mathcal{N}}}\left(\hat{a}^{\dagger}\hat{J}_{-}+\hat{a}\hat{J}_{+}\right)  \\
    &  + \frac{\gamma_{+}}{\sqrt{\mathcal{N}}}\left(\hat{a}^{\dagger}\hat{J}_{+}+\hat{a}\hat{J}_{-}\right), \nonumber
\end{align}
where $\hat{a}^{\dagger}$ ($\hat{a}$) is the bosonic creation (annihilation) operator of the field mode,  $\hat{J}_{x,y,z}=(1/2)\sum_{k=1}^{\mathcal{N}}\hat{\sigma}_{x,y,z}^{(k)}$ represent the collective pseudospin operators with $\hat{\sigma}_{x,y,z}^{(k)}$ being the Pauli matrices associated with each atom, $\hat{J}_{\pm}=\hat{J}_{x}\pm i\hat{J}_{y}$, and $\gamma_{-}$ ($\gamma_{+}$) is the coupling strength associated with the corotating (counterrotating) terms. The isotropic Dicke Hamiltonian has a single coupling strength $\gamma$, which means $\delta=1$ when we write $\gamma_{-}=\gamma$ and $\gamma_{+}= \delta \gamma$. 

The Dicke Hamiltonian commutes with the squared total pseudospin operator $\hat{\textbf{J}}^{2}=\hat{J}_{x}^{2}+\hat{J}_{y}^{2}+\hat{J}_{z}^{2}$, whose eigenvalues $j(j+1)$ identify invariant subspaces. We use the symmetric subspace, $j=\mathcal{N}/2$, that includes the ground state. The Hamiltonian also commutes with the parity operator $\hat{\Pi} = \text{exp}[i\pi(\hat{a}^{\dagger}\hat{a}+\hat{J}_{z}+j\hat{\mathbb{I}})]$. For $\gamma<\gamma_{\text{c},\delta}=\sqrt{\omega\omega_{0}}/(1+\delta)$, the system is in the normal phase and for $\gamma>\gamma_{\text{c},\delta}$, it is in the superradiant phase~\cite{Hioe1973,Hepp1973a,Hepp1973b,Carmichael1973,Wang1973}.

The Markovian Lindblad master equation for the open Dicke model is 
\begin{equation}
    \label{eq:DickeLiouvillian}
    \frac{d\hat{\rho}}{dt} = \hat{\mathcal{L}}_{\text{D}}\hat{\rho} = -i[\hat{H}_{\text{D}},\hat{\rho}] + \kappa\left(2\hat{a}\hat{\rho}\hat{a}^{\dagger} - \{\hat{a}^{\dagger}\hat{a},\hat{\rho}\}\right),
\end{equation}
where $\hat{\rho}$ is the density matrix operator of the system and $\kappa$ is the cavity decay coupling associated with photon leakage. The Dicke Liouvillian, $\hat{\mathcal{L}}_{\text{D}}$, commutes with the parity superoperator $\hat{\mathcal{P}}\hat{\rho}=\hat{\Pi}\hat{\rho}\hat{\Pi}^{\dagger}$ \cite{Buca2012,Albert2014,Lieu2020}. Analogously to the isolated system, the open system shows a dissipative quantum phase transition at the critical value~\cite{Dimer2007,Larson2017,Gutierrez2018,Boneberg2022,Lyu2024}
\begin{equation}
    \label{eq:DissipativeQPT}
    \gamma_{\text{c},\delta}^{\text{os}} = \frac{\gamma_{\text{c},\delta}}{1-\delta}\sqrt{1+\delta^{2}-2\delta\sqrt{1-\frac{(1-\delta^{2})^{2}\kappa^{2}}{4\delta^{2}\omega^{2}}}}.
\end{equation}
To perform the spectral analysis, we use the positive parity sector of the Dicke Liouvillian.

{\it Classical Dicke model---}
The classical limit of the isolated Dicke model is obtained by taking the expectation value of $\hat{H}_{\text{D}}$ under Glauber $|\alpha\rangle = e^{-|\alpha|^{2}/2}e^{\alpha\hat{a}^{\dagger}}|0\rangle$ and Bloch $|\beta\rangle = (1 + |\beta|^{2})^{-j}e^{\beta\hat{J}_{+}}|j,-j\rangle$ coherent states, where $|0\rangle$ is the vacuum Fock state and $|j,-j\rangle$ is the angular momentum state with all atoms in their ground state. The parameters $\alpha,\beta\in\mathbb{C}$ can be expressed in terms of classical variables. We associate position and momentum variables $(q,p)$ with the bosonic parameter $\alpha=\sqrt{j/2}(q+ip)$ and the Bloch sphere variables $(J_{x},J_{y},J_{z})$ with the collective pseudospin operators $\langle\beta|\hat{J}_{x,y,z}|\beta\rangle=jJ_{x,y,z}$, so the classical Hamiltonian scaled by the system size $j$ becomes 
\begin{align}
    \label{eq:ClassicalDickeHamiltonian}
    h_{\text{D}}(q,p,J_{x,y,z}) = & \; \frac{\omega}{2}\left(q^{2}+p^{2}\right) + \omega_{0}J_{z}  \\
    & + (\gamma_{-}+\gamma_{+})qJ_{x} - (\gamma_{-}-\gamma_{+})pJ_{y}. \nonumber
\end{align}
Both isotropic and anisotropic Hamiltonians display regular or chaotic behavior depending on the Hamiltonian parameters and excitation energies~\cite{Deaguiar1992,Bastarrachea2014b,Chavez2016}.

The classical limit of the open Dicke model is derived from the quantum evolution, $\frac{d}{dt}\hat{O} = \hat{\mathcal{L}}_{\text{D}}^{\dagger}\hat{O}$, of a given observable $\hat{O}$ and its expectation value $\langle \hat{O} \rangle = \text{Tr}(\hat{\rho}\hat{O})$, and by considering decoupled expectation values, $\langle\hat{O}_{1}\hat{O}_{2}\rangle \approx \langle\hat{O}_{1}\rangle\langle\hat{O}_{2}\rangle$. After scaling the expectation values of the following operators as $a = \langle \hat{a} \rangle / \sqrt{j} = \alpha / \sqrt{j} = (q + ip) / \sqrt{2}$ and $J_{x,y,z} = \langle \hat{J}_{x,y,z} \rangle / j$, we obtain 
\begin{align}
    \label{eq:qOpenDicke}
    \dot{q} & = -\kappa q + \omega p - (\gamma_{-}-\gamma_{+})J_{y}, \\
    \label{eq:pOpenDicke}
    \dot{p} & = -\kappa p - \omega q - (\gamma_{-}+\gamma_{+})J_{x}, \\
    \label{eq:JxOpenDicke}
    \dot{J}_{x} & = - \omega_{0} J_{y} - (\gamma_{-}-\gamma_{+})pJ_{z}, \\
    \label{eq:JyOpenDicke}
    \dot{J}_{y} & = \omega_{0} J_{x} - (\gamma_{-}+\gamma_{+})qJ_{z}, \\
    \label{eq:JzOpenDicke}
    \dot{J}_{z} & = (\gamma_{-}+\gamma_{+})qJ_{y} + (\gamma_{-}-\gamma_{+})pJ_{x}.
\end{align}

An exhaustive study of the classical dynamics of the open Dicke model was performed in Ref.~\cite{Stitely2020}, where a phase diagram of classical behavior as a function of the coupling strengths $\gamma_{-}$ and $\gamma_{+}$ was presented. For large coupling strengths, regions of chaotic attractors are found for the anisotropic system, but the isotropic case only shows simple attractors. The route to chaos is the usual infinite sequence of period-doubling bifurcations~\cite{Guckenheimer1983book,Shivamoggi2014book}. For small coupling strengths, the entire classical dynamics collapses to stable sinks~\cite{Stitely2020}. Aligned with this study, the analysis of the classical dynamics of the two-photon Dicke model with photon dissipation~\cite{Li2022} also required anisotropy for revealing chaotic attractors. A way to ensure chaotic motion in the isotropic two-photon Dicke model, but not in the standard Dicke model, is by including both photon and atomic dissipation~\cite{Li2024}. In this case, in addition to the infinite period-doubling bifurcation, two alternative routes to chaos exist, intermittency and quasiperiodicity. Other variations of the Dicke model have also been the subject of dissipative chaos~\cite{Vivek2024ARXIV}.

\begin{figure}[!t]
\centering
\includegraphics[width=0.95\columnwidth]{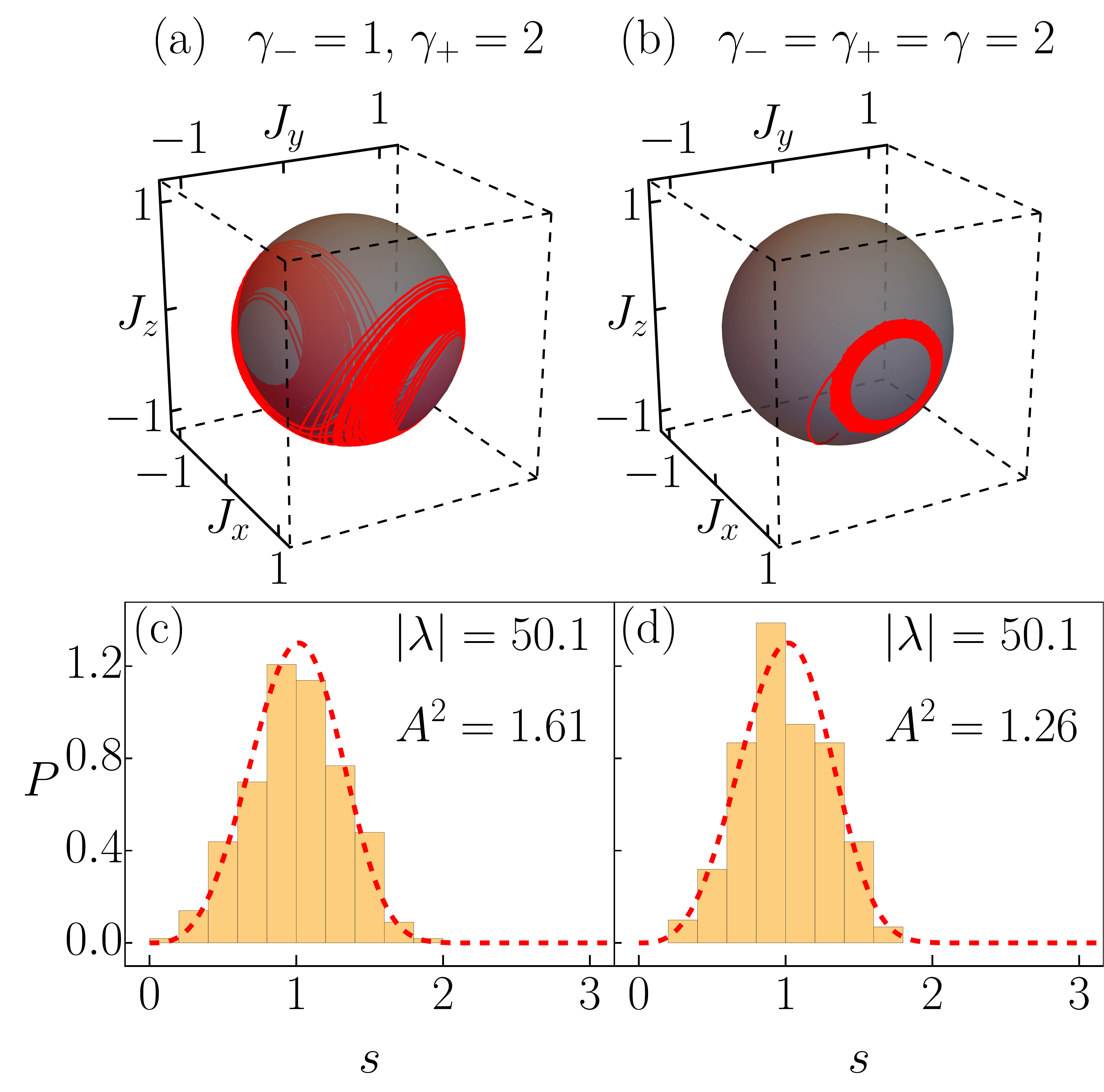}
\caption{Evolution on the Bloch sphere of the classical open Dicke model for an initial condition near the South Pole for the (a) anisotropic and (b) isotropic system.  Spacing distribution of the complex eigenvalues of the Dicke Liouvillian for the (c) anisotropic and (d) isotropic case. A chaotic attractor is seen in panel (a) and a limit cycle in panel (b), while both distributions in panels (c) and (d) follow the GinUE distribution, as confirmed also with the values of the Anderson-Darling statistical test $A^2$. Initial condition in panels (a)-(b): $(q_{0},p_{0},J_{x,y,z}^{0})=(0,0,0.001,0,-1)$. In panels (c)-(d): Unfolded eigenvalues contained in the windows centered at $|\lambda|=50.1$ for $j=1$. System parameters: $\omega=\omega_{0}=\kappa=1$.
}
\label{fig:BlochSphere}
\end{figure}

In Figs.~\ref{fig:BlochSphere}(a) and~\ref{fig:BlochSphere}(b), we show the classical evolution on the Bloch sphere for an initial condition near the South Pole for the open Dicke model in Eqs.~\eqref{eq:qOpenDicke}-\eqref{eq:JzOpenDicke} with large coupling strengths. A chaotic attractor appears for the anisotropic system in Fig.~\ref{fig:BlochSphere}(a), but a limit cycle is seen for the isotropic case in Fig.~\ref{fig:BlochSphere}(b). This contrasts with the isolated model, which is chaotic for the system parameters  in both panels. It is intriguing that dissipation erases any trace of the chaotic behavior present in the isolated isotropic model, but chaos survives for a range of system parameters in the open anisotropic model.

{\it Level statistics---}
Just as in the case of the isolated systems, the complex eigenvalues $\lambda$ of the Liouvillian need to be unfolded for the analysis of spectral correlations~\cite{Markum1999,Akemann2019,Hamazaki2020}. According to the GHS conjecture, if the dissipative systems show simple attractors in the classical dynamics, their level spacing distribution for the unfolded spectra is the 2D Poisson distribution~\cite{Grobe1988,Akemann2019},
\begin{equation}
    \label{eq:2DPDistribution}
    P_{\text{2DP}}(s) = \frac{\pi}{2}s\,e^{-\pi s^{2}/4},
\end{equation}
 while for systems that show chaotic attractors in the classical dynamics, $P(s)$  follows the Ginibre unitary ensemble (GinUE) distribution~\cite{Ginibre1965,Grobe1988,Akemann2019,Hamazaki2020}
\begin{equation}
    \label{eq:GinUEDistribution}
    P_{\text{GinUE}}(s) = \prod_{k=1}^{\infty}\frac{\Gamma(1+k,s^{2})}{k!}\sum_{k'=1}^{\infty}\frac{2s^{2k'+1}e^{-s^{2}}}{\Gamma(1+k',s^{2})} ,
\end{equation}
where $\Gamma(k,z) = \int_{z}^{\infty}dt\,t^{k-1}e^{-t}$ is the incomplete Gamma function. The first moment of the GinUE distribution is $\bar{s} = \int_{0}^{\infty} ds\,s\,P_{\text{GinUE}}(s)\approx1.1429$, so when comparing the distribution with numerical results, the scaling $\widetilde{P}_{\text{GinUE}}(s) = \bar{s}P_{\text{GinUE}}(\bar{s}s)$ is done to ensure that the first moment is unity, $\int_{0}^{\infty} ds\,s\,\widetilde{P}_{\text{GinUE}}(s) = 1$. The degree of level repulsion is determined by the limit $s\to0$ of the distributions above, which gives $P_{\beta}(s) \propto s^{\beta}$. Linear (cubic) level repulsion, $\beta=1$ ($\beta=3$), indicates a regular (chaotic) dissipative quantum system~\cite{HaakeBook,Grobe1988,Grobe1989}.

{\it Breakdown of the GHS conjecture---}
In Figs.~\ref{fig:BlochSphere}(c) and~\ref{fig:BlochSphere}(d), we show the spacing distribution for the eigenvalues of the Dicke Liouvillian in Eq.~\eqref{eq:DickeLiouvillian}.  The distributions for both the anisotropic [Fig.~\ref{fig:BlochSphere}(c)] and the isotropic [Fig.~\ref{fig:BlochSphere}(d)] model agree with the GinUE distribution, which contrasts with the results for the classical limit, where a chaotic attractor is only seen for the anisotropic case [Fig.~\ref{fig:BlochSphere}(a)]. The fact that the classical open isotropic system develops simple attractors [Fig.~\ref{fig:BlochSphere}(b)], while its quantum counterpart can exhibit GinUE spectral correlations [Fig.~\ref{fig:BlochSphere}(d)] indicates the failure of the GHS conjecture for the Dicke model. 

In Figs.~\ref{fig:BlochSphere}(c) and~\ref{fig:BlochSphere}(d), we further corroborate the agreement between the numerical data and the analytical GinUE distribution using the Anderson-Darling statistical test $A^2$~\cite{Anderson1952}. Agreement implies $A^2<2.5$ (see  Supplemental Material~\cite{footSM} for more details and additional figures). 

\begin{figure}[!t]
\centering
\includegraphics[width=0.95\columnwidth]{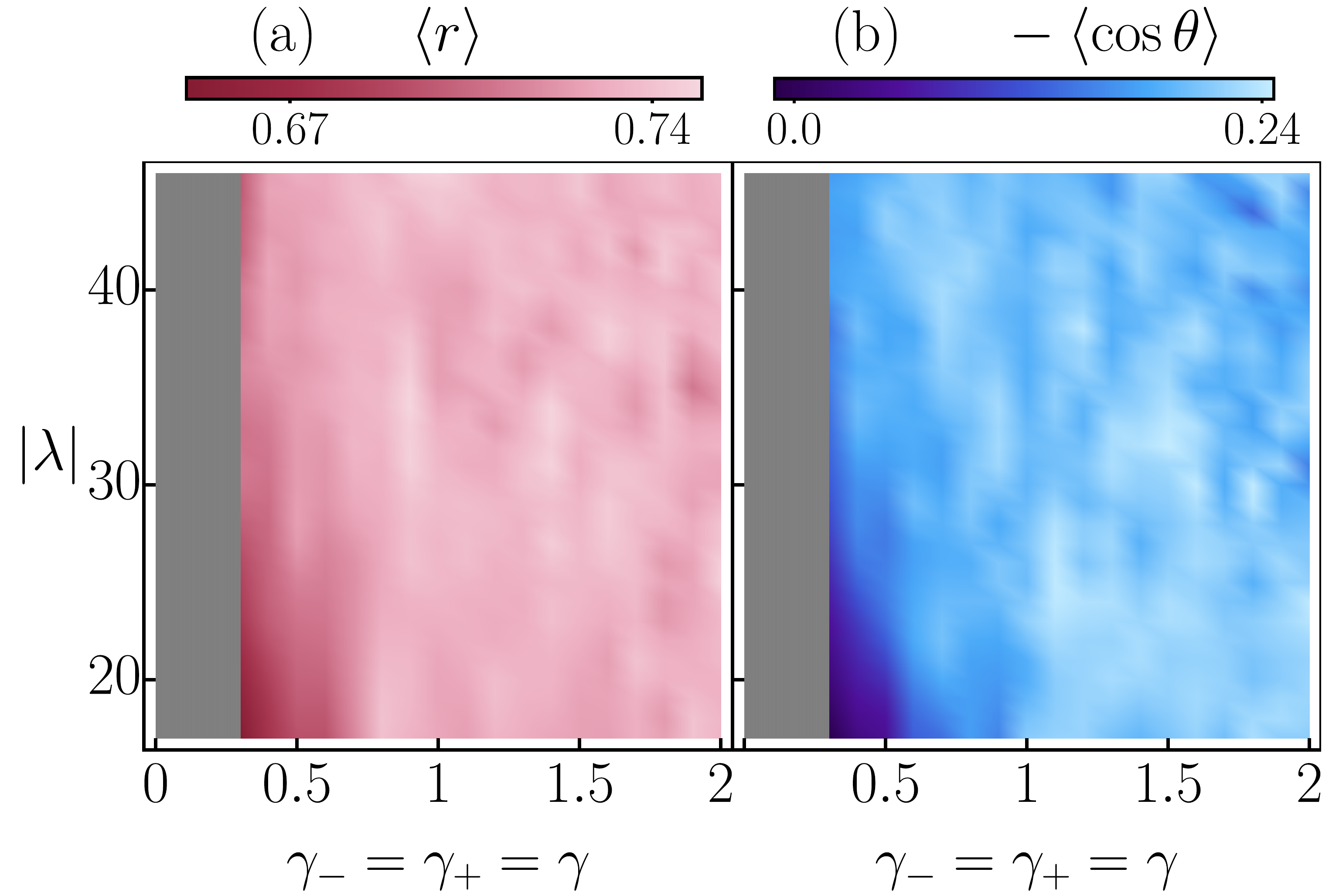}
\caption{Analysis of  the complex spectral ratio in Eq.~\eqref{eq:ComplexRatio} for the open isotropic Dicke model. (a) Average $\langle r \rangle$  and (b) average $-\langle \cos\theta \rangle$ as a function of the absolute values of the converged Liouvillian eigenvalues $|\lambda|$ and the coupling strength $\gamma$. The average is performed with moving windows of approximately 3000 (1000) eigenvalues for low (high) coupling strengths. Dark (light) colors indicate absence (presence) of correlated levels. The gray solid region represents the coupling values for which neither the 2DP Poisson nor the GinUE spacing distribution holds. System parameters: $\omega=\omega_{0}=\kappa=1$ and $j=2$.
}
\label{fig:IsotropicOpenDicke}
\end{figure}

Figure~\ref{fig:IsotropicOpenDicke} extends the analysis of Fig.~\ref{fig:BlochSphere}(d). While the classical open isotropic Dicke model only presents simple attractors~\cite{Stitely2020}, our results in Fig.~\ref{fig:IsotropicOpenDicke} show that the GinUE spectral correlations predominate for the quantum counterpart. In Fig.~\ref{fig:IsotropicOpenDicke}, to avoid the unfolding procedure, similarly to what is done for isolated systems~\cite{Oganesyan2007,Atas2013}, the analysis of level statistics of complex eigenvalues $\lambda_{k}$ uses the complex spectral ratio~\cite{Sa2020}
\begin{equation}
    \label{eq:ComplexRatio}
    Z_{k}=r_{k}e^{i\theta_{k}}=\frac{\lambda_{k}^{\text{NN}}-\lambda_{k}}{\lambda_{k}^{\text{NNN}}-\lambda_{k}} ,
\end{equation}
where NN stands for nearest neighbor and NNN for next-to-nearest neighbor. There are two limits associated with the absolute value $r_{k}=|Z_{k}|$ and the argument $\theta_{k}=\text{Arg}(Z_{k})$ of the complex ratio $Z_{k}$. For the 2D Poisson distribution, we have the average $\langle r\rangle_{\text{2DP}} = 2/3$  and $-\langle \cos\theta\rangle_{\text{2DP}} = 0$, while for the GinUE distribution, we have $\langle r\rangle_{\text{GinUE}} \approx 0.74$ and $-\langle \cos\theta\rangle_{\text{GinUE}} \approx 0.24$.

Figure~\ref{fig:IsotropicOpenDicke} provides density plots for the averages $\langle r\rangle$ [Fig.~\ref{fig:IsotropicOpenDicke}(a)] and $-\langle \cos\theta \rangle$ [Fig.~\ref{fig:IsotropicOpenDicke}(b)] as a function of $\gamma$ and the absolute values of the converged Liouvillian eigenvalues $|\lambda|$. For low coupling strengths, the low eigenvalues are mainly uncorrelated and tend to follow the 2D Poisson distribution (dark colors), while the high eigenvalues become correlated agreeing with the GinUE distribution (light colors). For large coupling strengths, the overall behavior is described by the GinUE distribution for low and high eigenvalues. In the gray region for $\gamma\to0$, the 2D Poisson and GinUE distributions do not hold, because the system is essentially composed of two harmonic oscillators with dissipation.

\begin{figure}[!t]
\centering
\includegraphics[width=0.95\columnwidth]{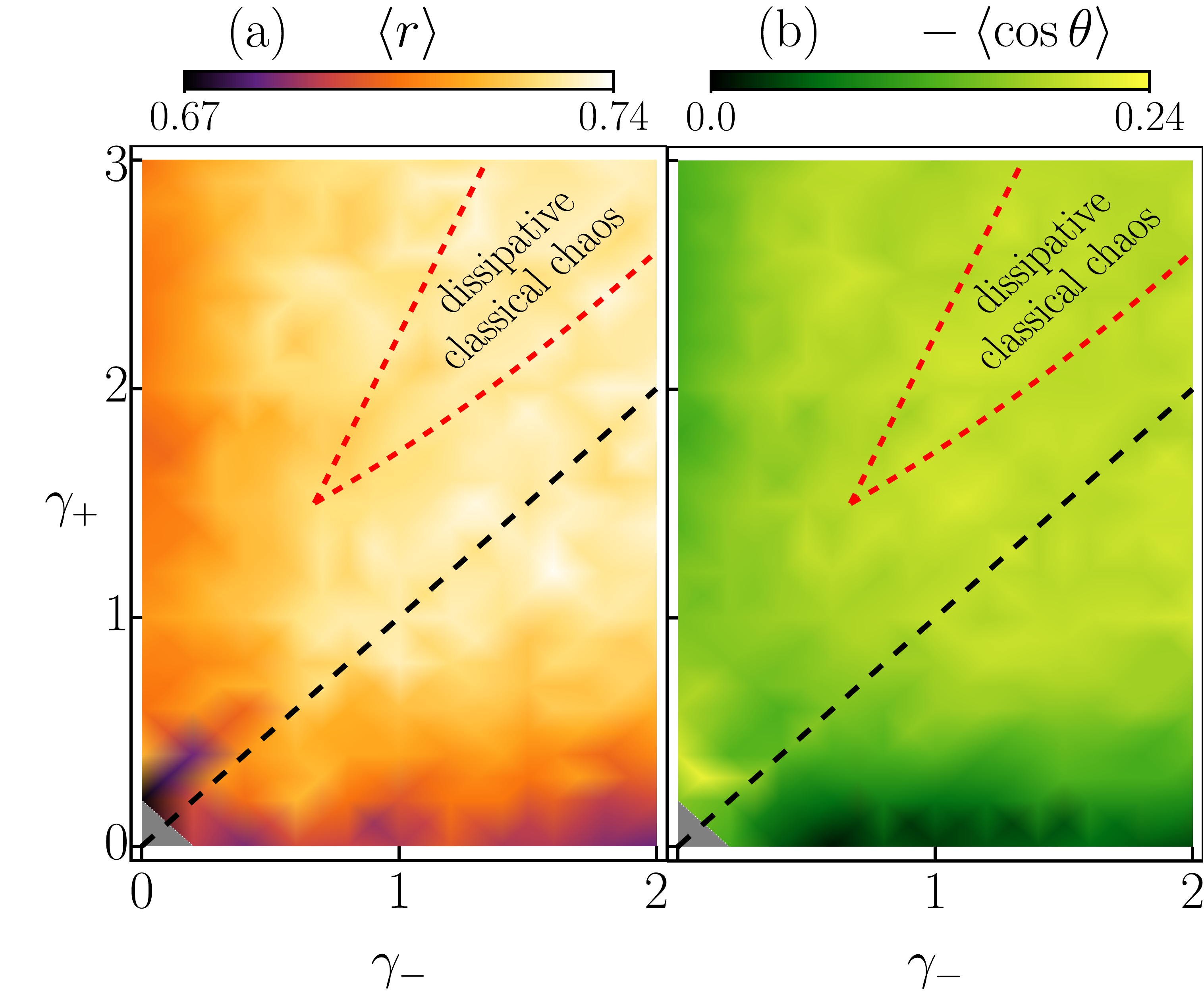}
\caption{Analysis of the complex spectral ratio in Eq.~\eqref{eq:ComplexRatio} for the open anisotropic Dicke model. (a) Average $\langle r \rangle$  and (b) average $-\langle \cos\theta \rangle$ as a function of the corotating ($\gamma_{-}$) and counterrotating ($\gamma_{+}$) coupling strengths. The average is performed over all converged eigenvalue absolute values $|\lambda|$ (about $10^4$). Dark (light) colors indicate absence (presence) of correlated levels. The red dotted lines indicate the region where chaotic attractors can be found in the classical dissipative dynamics of the anisotropic model. The diagonal black dashed line represents the isotropic Dicke model. The small gray triangle represents the coupling values for which neither the 2DP Poisson nor the GinUE spacing distribution holds. System parameters: $\omega=\omega_{0}=\kappa=1$ and $j=1$.
}
\label{fig:AnisotropicOpenDicke}
\end{figure}

{\it Breakdown of the GHS conjecture beyond the isotropic case---}
As we show next, the breakdown of the GHS conjecture is not exclusive to the isotropic case. In Fig.~\ref{fig:AnisotropicOpenDicke}, we depict $\langle r \rangle$ and  $-\langle \cos\theta \rangle$ averaged over all converged eigenvalue absolute values $|\lambda|$ as a function of the corotating ($\gamma_{-}$) and counterrotating ($\gamma_{+}$) coupling strengths. Corroborating the results in Ref.~\cite{Stitely2020}, we mark with red dotted lines the range of values of  $\gamma_{-}$ and $\gamma_{+}$ where chaotic attractors appear, which reiterates that dissipative classical chaos is only possible for this model if it is anisotropic and has large values of the coupling strengths. In contrast, the analysis of level statistics of the quantum dissipative model shows GinUE spectral correlations (light colors) for both the isotropic (black dashed line in Fig.~\ref{fig:AnisotropicOpenDicke}) and the anisotropic case. There is a broad range of coupling strengths outside the region of dissipative classical chaos, where the eigenvalues are correlated. The appearance of chaotic attractors is not a necessary condition for the onset of cubic level repulsion.

For completeness, we describe two special regions of system parameters in Fig.~\ref{fig:AnisotropicOpenDicke}. The case $\gamma_{+}=0$ corresponds to the Tavis-Cummings model~\cite{Tavis1968} with photon dissipation. The isolated classical Tavis-Cummings Hamiltonian is integrable~\cite{Bastarrachea2014b}, so unsurprisingly the open classical model for any value of $\gamma_{-}$ only shows simple attractors and the level statistics analysis of the quantum system indicates uncorrelated eigenvalues. The other region is the one for small values of $\gamma_{-}$, where the dominant terms of the system are counterrotating. In this case, we identify cubic level repulsion as  $\gamma_{+}$ grows, suggesting that the counterrotating terms play an important role in the generation of GinUE spectral correlations.

{\it Discussion---}
Despite the validity of the Bohigas-Giannoni-Schmit conjecture for the isolated Dicke model, we showed that the GHS conjecture fails for the open Dicke model. For a wide range of system parameters where the dissipative classical model does not exhibit chaos (absence of chaotic attractors), the dissipative quantum model can present GinUE spectral correlations. It remains to verify whether the breakdown of the GHS conjecture is exclusive to the Dicke model or extends to other interacting spin-boson systems and even beyond. This breakdown raises two important questions: (i) If chaotic motion in a dissipative classical system is not necessarily the source of GinUE spectral correlations in the quantum domain, then what is their origin? (ii) Could there be better quantum signatures of dissipative classical chaos than cubic level repulsion that are worth exploring, such as specific properties of dissipative out-of-time-ordered correlators~\cite{Garcia2024}?

Another fundamental question that emerged from our analysis concerns the classical Dicke model. It is not clear why there are regions of system parameters where the isolated model is chaotic, while the open system is not. We hope that the questions raised by our work will inspire new studies of the various aspects of quantum and classical dissipative systems.

{\it Acknowledgments---}
We acknowledge the support of the Computation Center - IIMAS, in particular to Adri\'an Chavesti. We also acknowledge the support of the Computation Center - ICN, in particular to Enrique Palacios, Luciano D\'iaz, and Eduardo Murrieta. D.V. acknowledges financial support from the postdoctoral fellowship program DGAPA-UNAM. L.F.S. is supported by the United States National Science Foundation (NSF, Grant No. DMR-1936006). P.B.-B. acknowledges support from the
PAPIIT-DGAPA Grant No. IG101324.

\bibliography{main}

\begin{thebibliography}{92}%
\makeatletter
\providecommand \@ifxundefined [1]{%
 \@ifx{#1\undefined}
}%
\providecommand \@ifnum [1]{%
 \ifnum #1\expandafter \@firstoftwo
 \else \expandafter \@secondoftwo
 \fi
}%
\providecommand \@ifx [1]{%
 \ifx #1\expandafter \@firstoftwo
 \else \expandafter \@secondoftwo
 \fi
}%
\providecommand \natexlab [1]{#1}%
\providecommand \enquote  [1]{``#1''}%
\providecommand \bibnamefont  [1]{#1}%
\providecommand \bibfnamefont [1]{#1}%
\providecommand \citenamefont [1]{#1}%
\providecommand \href@noop [0]{\@secondoftwo}%
\providecommand \href [0]{\begingroup \@sanitize@url \@href}%
\providecommand \@href[1]{\@@startlink{#1}\@@href}%
\providecommand \@@href[1]{\endgroup#1\@@endlink}%
\providecommand \@sanitize@url [0]{\catcode `\\12\catcode `\$12\catcode
  `\&12\catcode `\#12\catcode `\^12\catcode `\_12\catcode `\%12\relax}%
\providecommand \@@startlink[1]{}%
\providecommand \@@endlink[0]{}%
\providecommand \url  [0]{\begingroup\@sanitize@url \@url }%
\providecommand \@url [1]{\endgroup\@href {#1}{\urlprefix }}%
\providecommand \urlprefix  [0]{URL }%
\providecommand \Eprint [0]{\href }%
\providecommand \doibase [0]{http://dx.doi.org/}%
\providecommand \selectlanguage [0]{\@gobble}%
\providecommand \bibinfo  [0]{\@secondoftwo}%
\providecommand \bibfield  [0]{\@secondoftwo}%
\providecommand \translation [1]{[#1]}%
\providecommand \BibitemOpen [0]{}%
\providecommand \bibitemStop [0]{}%
\providecommand \bibitemNoStop [0]{.\EOS\space}%
\providecommand \EOS [0]{\spacefactor3000\relax}%
\providecommand \BibitemShut  [1]{\csname bibitem#1\endcsname}%
\let\auto@bib@innerbib\@empty
\bibitem [{\citenamefont {Ott}(2002)}]{OttBook}%
  \BibitemOpen
  \bibfield  {author} {\bibinfo {author} {\bibfnamefont {Edward}\ \bibnamefont
  {Ott}},\ }\href@noop {} {\emph {\bibinfo {title} {Chaos in Dynamical
  Systems}}}\ (\bibinfo  {publisher} {Cambridge University Press},\ \bibinfo
  {address} {Cambridge, UK},\ \bibinfo {year} {2002})\BibitemShut {NoStop}%
\bibitem [{\citenamefont {Gutzwiller}(1990)}]{Gutzwiller1990book}%
  \BibitemOpen
  \bibfield  {author} {\bibinfo {author} {\bibfnamefont {M.~C.}\ \bibnamefont
  {Gutzwiller}},\ }\href@noop {} {\emph {\bibinfo {title} {Chaos in classical
  and quantum mechanics}}}\ (\bibinfo  {publisher} {Springer-Verlag},\ \bibinfo
  {address} {New York},\ \bibinfo {year} {1990})\BibitemShut {NoStop}%
\bibitem [{\citenamefont {Bohigas}\ \emph {et~al.}(1984)\citenamefont
  {Bohigas}, \citenamefont {Giannoni},\ and\ \citenamefont
  {Schmit}}]{Bohigas1984}%
  \BibitemOpen
  \bibfield  {author} {\bibinfo {author} {\bibfnamefont {O.}~\bibnamefont
  {Bohigas}}, \bibinfo {author} {\bibfnamefont {M.~J.}\ \bibnamefont
  {Giannoni}}, \ and\ \bibinfo {author} {\bibfnamefont {C.}~\bibnamefont
  {Schmit}},\ }\bibfield  {title} {\enquote {\bibinfo {title} {Characterization
  of chaotic quantum spectra and universality of level fluctuation laws},}\
  }\href {\doibase 10.1103/PhysRevLett.52.1} {\bibfield  {journal} {\bibinfo
  {journal} {Phys. Rev. Lett.}\ }\textbf {\bibinfo {volume} {52}},\ \bibinfo
  {pages} {1--4} (\bibinfo {year} {1984})}\BibitemShut {NoStop}%
\bibitem [{\citenamefont {Casati}\ \emph {et~al.}(1980)\citenamefont {Casati},
  \citenamefont {Valz-Gris},\ and\ \citenamefont {Guarnieri}}]{Casati1980}%
  \BibitemOpen
  \bibfield  {author} {\bibinfo {author} {\bibfnamefont {G.}~\bibnamefont
  {Casati}}, \bibinfo {author} {\bibfnamefont {F.}~\bibnamefont {Valz-Gris}}, \
  and\ \bibinfo {author} {\bibfnamefont {I.}~\bibnamefont {Guarnieri}},\
  }\bibfield  {title} {\enquote {\bibinfo {title} {On the connection between
  quantization of nonintegrable systems and statistical theory of spectra},}\
  }\href {\doibase 10.1007/BF02798790} {\bibfield  {journal} {\bibinfo
  {journal} {Lett. Nuovo Cimento}\ }\textbf {\bibinfo {volume} {28}},\ \bibinfo
  {pages} {279--282} (\bibinfo {year} {1980})}\BibitemShut {NoStop}%
\bibitem [{\citenamefont {Mehta}(1991)}]{MehtaBook}%
  \BibitemOpen
  \bibfield  {author} {\bibinfo {author} {\bibfnamefont {M.~L.}\ \bibnamefont
  {Mehta}},\ }\href@noop {} {\emph {\bibinfo {title} {Random Matrices}}}\
  (\bibinfo  {publisher} {Academic Press},\ \bibinfo {address} {Boston, MA},\
  \bibinfo {year} {1991})\BibitemShut {NoStop}%
\bibitem [{\citenamefont {Haake}(1991)}]{HaakeBook}%
  \BibitemOpen
  \bibfield  {author} {\bibinfo {author} {\bibfnamefont {Fritz}\ \bibnamefont
  {Haake}},\ }\href@noop {} {\emph {\bibinfo {title} {Quantum Signatures of
  Chaos}}}\ (\bibinfo  {publisher} {Springer-Verlag},\ \bibinfo {address}
  {Berlin},\ \bibinfo {year} {1991})\BibitemShut {NoStop}%
\bibitem [{\citenamefont {Wimberger}(2014)}]{WimbergerBook}%
  \BibitemOpen
  \bibfield  {author} {\bibinfo {author} {\bibfnamefont {Sandro}\ \bibnamefont
  {Wimberger}},\ }\href@noop {} {\emph {\bibinfo {title} {Nonlinear Dynamics
  and Quantum Chaos}}}\ (\bibinfo  {publisher} {Springer International
  Publishing},\ \bibinfo {address} {Switzerland},\ \bibinfo {year}
  {2014})\BibitemShut {NoStop}%
\bibitem [{\citenamefont {Guhr}\ \emph {et~al.}(1998)\citenamefont {Guhr},
  \citenamefont {M\"uller-Groeling},\ and\ \citenamefont
  {Weidenm\"uller}}]{Guhr1998}%
  \BibitemOpen
  \bibfield  {author} {\bibinfo {author} {\bibfnamefont {T.}~\bibnamefont
  {Guhr}}, \bibinfo {author} {\bibfnamefont {A.}~\bibnamefont
  {M\"uller-Groeling}}, \ and\ \bibinfo {author} {\bibfnamefont {H.~A.}\
  \bibnamefont {Weidenm\"uller}},\ }\bibfield  {title} {\enquote {\bibinfo
  {title} {Random matrix theories in quantum physics: Common concepts},}\
  }\href {\doibase https://doi.org/10.1016/S0370-1573(97)00088-4} {\bibfield
  {journal} {\bibinfo  {journal} {Phys. Rep.}\ }\textbf {\bibinfo {volume}
  {299}},\ \bibinfo {pages} {189} (\bibinfo {year} {1998})}\BibitemShut
  {NoStop}%
\bibitem [{\citenamefont {Berry}(1977)}]{Berry1977JPA}%
  \BibitemOpen
  \bibfield  {author} {\bibinfo {author} {\bibfnamefont {M.~V.}\ \bibnamefont
  {Berry}},\ }\bibfield  {title} {\enquote {\bibinfo {title} {Regular and
  irregular semiclassical wavefunctions},}\ }\href {\doibase
  10.1088/0305-4470/10/12/016} {\bibfield  {journal} {\bibinfo  {journal} {J.
  Phys. A}\ }\textbf {\bibinfo {volume} {10}},\ \bibinfo {pages} {2083--2091}
  (\bibinfo {year} {1977})}\BibitemShut {NoStop}%
\bibitem [{\citenamefont {Borgonovi}\ \emph {et~al.}(2006)\citenamefont
  {Borgonovi}, \citenamefont {Celardo}, \citenamefont {Musesti}, \citenamefont
  {Trasarti-Battistoni},\ and\ \citenamefont {Vachal}}]{Borgonovi2006}%
  \BibitemOpen
  \bibfield  {author} {\bibinfo {author} {\bibfnamefont {F.}~\bibnamefont
  {Borgonovi}}, \bibinfo {author} {\bibfnamefont {G.~L.}\ \bibnamefont
  {Celardo}}, \bibinfo {author} {\bibfnamefont {A.}~\bibnamefont {Musesti}},
  \bibinfo {author} {\bibfnamefont {R.}~\bibnamefont {Trasarti-Battistoni}}, \
  and\ \bibinfo {author} {\bibfnamefont {P.}~\bibnamefont {Vachal}},\
  }\bibfield  {title} {\enquote {\bibinfo {title} {Topological nonconnectivity
  threshold in long-range spin systems},}\ }\href {\doibase
  10.1103/PhysRevE.73.026116} {\bibfield  {journal} {\bibinfo  {journal} {Phys.
  Rev. E}\ }\textbf {\bibinfo {volume} {73}},\ \bibinfo {pages} {026116}
  (\bibinfo {year} {2006})}\BibitemShut {NoStop}%
\bibitem [{\citenamefont {Berry}\ \emph {et~al.}(1977)\citenamefont {Berry},
  \citenamefont {Tabor},\ and\ \citenamefont {Ziman}}]{Berry1977}%
  \BibitemOpen
  \bibfield  {author} {\bibinfo {author} {\bibfnamefont {Michael~Victor}\
  \bibnamefont {Berry}}, \bibinfo {author} {\bibfnamefont {M.}~\bibnamefont
  {Tabor}}, \ and\ \bibinfo {author} {\bibfnamefont {John~Michael}\
  \bibnamefont {Ziman}},\ }\bibfield  {title} {\enquote {\bibinfo {title}
  {Level clustering in the regular spectrum},}\ }\href {\doibase
  10.1098/rspa.1977.0140} {\bibfield  {journal} {\bibinfo  {journal} {Proc.
  Roy. Soc. London. A. Math. Phys. Sci.}\ }\textbf {\bibinfo {volume} {356}},\
  \bibinfo {pages} {375--394} (\bibinfo {year} {1977})}\BibitemShut {NoStop}%
\bibitem [{\citenamefont {Seligman}\ \emph {et~al.}(1984)\citenamefont
  {Seligman}, \citenamefont {Verbaarschot},\ and\ \citenamefont
  {Zirnbauer}}]{Seligman1984}%
  \BibitemOpen
  \bibfield  {author} {\bibinfo {author} {\bibfnamefont {T.~H.}\ \bibnamefont
  {Seligman}}, \bibinfo {author} {\bibfnamefont {J.~J.~M.}\ \bibnamefont
  {Verbaarschot}}, \ and\ \bibinfo {author} {\bibfnamefont {M.~R.}\
  \bibnamefont {Zirnbauer}},\ }\bibfield  {title} {\enquote {\bibinfo {title}
  {Quantum spectra and transition from regular to chaotic classical motion},}\
  }\href {\doibase 10.1103/PhysRevLett.53.215} {\bibfield  {journal} {\bibinfo
  {journal} {Phys. Rev. Lett.}\ }\textbf {\bibinfo {volume} {53}},\ \bibinfo
  {pages} {215--217} (\bibinfo {year} {1984})}\BibitemShut {NoStop}%
\bibitem [{\citenamefont {Seligman}\ \emph {et~al.}(1985)\citenamefont
  {Seligman}, \citenamefont {Verbaarschot},\ and\ \citenamefont
  {Zirnbauer}}]{Seligman1985}%
  \BibitemOpen
  \bibfield  {author} {\bibinfo {author} {\bibfnamefont {T~H}\ \bibnamefont
  {Seligman}}, \bibinfo {author} {\bibfnamefont {J~J~M}\ \bibnamefont
  {Verbaarschot}}, \ and\ \bibinfo {author} {\bibfnamefont {M~R}\ \bibnamefont
  {Zirnbauer}},\ }\bibfield  {title} {\enquote {\bibinfo {title} {Spectral
  fluctuation properties of {H}amiltonian systems: the transition region
  between order and chaos},}\ }\href {\doibase 10.1088/0305-4470/18/14/026}
  {\bibfield  {journal} {\bibinfo  {journal} {J. Phys. A: Math. Gen.}\ }\textbf
  {\bibinfo {volume} {18}},\ \bibinfo {pages} {2751} (\bibinfo {year}
  {1985})}\BibitemShut {NoStop}%
\bibitem [{\citenamefont {Friedrich}\ and\ \citenamefont
  {Wintgen}(1989)}]{Friedrich1989}%
  \BibitemOpen
  \bibfield  {author} {\bibinfo {author} {\bibfnamefont {Harald}\ \bibnamefont
  {Friedrich}}\ and\ \bibinfo {author} {\bibfnamefont {Dieter}\ \bibnamefont
  {Wintgen}},\ }\bibfield  {title} {\enquote {\bibinfo {title} {The hydrogen
  atom in a uniform magnetic field — an example of chaos},}\ }\href {\doibase
  10.1016/0370-1573(89)90121-X} {\bibfield  {journal} {\bibinfo  {journal}
  {Phys. Rep.}\ }\textbf {\bibinfo {volume} {183}},\ \bibinfo {pages} {37--79}
  (\bibinfo {year} {1989})}\BibitemShut {NoStop}%
\bibitem [{\citenamefont {Haake}\ \emph {et~al.}(1987)\citenamefont {Haake},
  \citenamefont {Ku\'s},\ and\ \citenamefont {Scharf}}]{Haake1987}%
  \BibitemOpen
  \bibfield  {author} {\bibinfo {author} {\bibfnamefont {F.}~\bibnamefont
  {Haake}}, \bibinfo {author} {\bibfnamefont {M.}~\bibnamefont {Ku\'s}}, \ and\
  \bibinfo {author} {\bibfnamefont {R.}~\bibnamefont {Scharf}},\ }\bibfield
  {title} {\enquote {\bibinfo {title} {Classical and quantum chaos for a kicked
  top},}\ }\href {\doibase 10.1007/BF01303727} {\bibfield  {journal} {\bibinfo
  {journal} {Z. Phys. B Cond. Matt.}\ }\textbf {\bibinfo {volume} {65}},\
  \bibinfo {pages} {381--365} (\bibinfo {year} {1987})}\BibitemShut {NoStop}%
\bibitem [{\citenamefont {Chaudhury}\ \emph {et~al.}(2009)\citenamefont
  {Chaudhury}, \citenamefont {Smith}, \citenamefont {Anderson}, \citenamefont
  {Ghose},\ and\ \citenamefont {Jessen}}]{Chaudhury2009}%
  \BibitemOpen
  \bibfield  {author} {\bibinfo {author} {\bibfnamefont {S.}~\bibnamefont
  {Chaudhury}}, \bibinfo {author} {\bibfnamefont {A.}~\bibnamefont {Smith}},
  \bibinfo {author} {\bibfnamefont {B.~E.}\ \bibnamefont {Anderson}}, \bibinfo
  {author} {\bibfnamefont {S.}~\bibnamefont {Ghose}}, \ and\ \bibinfo {author}
  {\bibfnamefont {P.~S.}\ \bibnamefont {Jessen}},\ }\bibfield  {title}
  {\enquote {\bibinfo {title} {Quantum signatures of chaos in a kicked top},}\
  }\href {https://doi.org/10.1038/nature08396} {\bibfield  {journal} {\bibinfo
  {journal} {Nature}\ }\textbf {\bibinfo {volume} {461}},\ \bibinfo {pages}
  {768--771} (\bibinfo {year} {2009})}\BibitemShut {NoStop}%
\bibitem [{\citenamefont {Izrailev}(1990)}]{Izrailev1990}%
  \BibitemOpen
  \bibfield  {author} {\bibinfo {author} {\bibfnamefont {F.~M.}\ \bibnamefont
  {Izrailev}},\ }\bibfield  {title} {\enquote {\bibinfo {title} {Simple models
  of quantum chaos: Spectrum and eigenfunctions},}\ }\href {\doibase
  10.1016/0370-1573(90)90067-C} {\bibfield  {journal} {\bibinfo  {journal}
  {Phys. Rep.}\ }\textbf {\bibinfo {volume} {196}},\ \bibinfo {pages}
  {299--392} (\bibinfo {year} {1990})}\BibitemShut {NoStop}%
\bibitem [{\citenamefont {Santhanam}\ \emph {et~al.}(2022)\citenamefont
  {Santhanam}, \citenamefont {Paul},\ and\ \citenamefont
  {Kannan}}]{Santhanam2022}%
  \BibitemOpen
  \bibfield  {author} {\bibinfo {author} {\bibfnamefont {M.~S.}\ \bibnamefont
  {Santhanam}}, \bibinfo {author} {\bibfnamefont {Sanku}\ \bibnamefont {Paul}},
  \ and\ \bibinfo {author} {\bibfnamefont {J.~Bharathi}\ \bibnamefont
  {Kannan}},\ }\bibfield  {title} {\enquote {\bibinfo {title} {Quantum kicked
  rotor and its variants: Chaos, localization and beyond},}\ }\href {\doibase
  10.1016/j.physrep.2022.01.002} {\bibfield  {journal} {\bibinfo  {journal}
  {Phys. Rep.}\ }\textbf {\bibinfo {volume} {956}},\ \bibinfo {pages} {1--87}
  (\bibinfo {year} {2022})}\BibitemShut {NoStop}%
\bibitem [{\citenamefont {Wang}\ \emph {et~al.}(2022)\citenamefont {Wang},
  \citenamefont {Benenti}, \citenamefont {Casati},\ and\ \citenamefont
  {ge~Wang}}]{Wang2022}%
  \BibitemOpen
  \bibfield  {author} {\bibinfo {author} {\bibfnamefont {Jiaozi}\ \bibnamefont
  {Wang}}, \bibinfo {author} {\bibfnamefont {Giuliano}\ \bibnamefont
  {Benenti}}, \bibinfo {author} {\bibfnamefont {Giulio}\ \bibnamefont
  {Casati}}, \ and\ \bibinfo {author} {\bibfnamefont {Wen}\ \bibnamefont
  {ge~Wang}},\ }\bibfield  {title} {\enquote {\bibinfo {title} {Statistical and
  dynamical properties of the quantum triangle map},}\ }\href {\doibase
  10.1088/1751-8121/ac6a93} {\bibfield  {journal} {\bibinfo  {journal} {J.
  Phys. A: Math. Theor.}\ }\textbf {\bibinfo {volume} {55}},\ \bibinfo {pages}
  {234002} (\bibinfo {year} {2022})}\BibitemShut {NoStop}%
\bibitem [{\citenamefont {Hsu}\ and\ \citenamefont
  {Angle`s~d'Auriac}(1993)}]{Hsu1993}%
  \BibitemOpen
  \bibfield  {author} {\bibinfo {author} {\bibfnamefont {Theodore~C.}\
  \bibnamefont {Hsu}}\ and\ \bibinfo {author} {\bibfnamefont {J.~C.}\
  \bibnamefont {Angle`s~d'Auriac}},\ }\bibfield  {title} {\enquote {\bibinfo
  {title} {Level repulsion in integrable and almost-integrable quantum spin
  models},}\ }\href {\doibase 10.1103/PhysRevB.47.14291} {\bibfield  {journal}
  {\bibinfo  {journal} {Phys. Rev. B}\ }\textbf {\bibinfo {volume} {47}},\
  \bibinfo {pages} {14291--14296} (\bibinfo {year} {1993})}\BibitemShut
  {NoStop}%
\bibitem [{\citenamefont {Santos}\ \emph {et~al.}(2020)\citenamefont {Santos},
  \citenamefont {P{\'e}rez-Bernal},\ and\ \citenamefont
  {Torres-Herrera}}]{Santos2020}%
  \BibitemOpen
  \bibfield  {author} {\bibinfo {author} {\bibfnamefont {Lea~F.}\ \bibnamefont
  {Santos}}, \bibinfo {author} {\bibfnamefont {Francisco}\ \bibnamefont
  {P{\'e}rez-Bernal}}, \ and\ \bibinfo {author} {\bibfnamefont {E.~Jonathan}\
  \bibnamefont {Torres-Herrera}},\ }\bibfield  {title} {\enquote {\bibinfo
  {title} {Speck of chaos},}\ }\href {\doibase
  10.1103/PhysRevResearch.2.043034} {\bibfield  {journal} {\bibinfo  {journal}
  {Phys. Rev. Res.}\ }\textbf {\bibinfo {volume} {2}},\ \bibinfo {pages}
  {043034} (\bibinfo {year} {2020})}\BibitemShut {NoStop}%
\bibitem [{\citenamefont {Casati}\ \emph {et~al.}(1985)\citenamefont {Casati},
  \citenamefont {Chirikov},\ and\ \citenamefont {Guarneri}}]{Casati1985}%
  \BibitemOpen
  \bibfield  {author} {\bibinfo {author} {\bibfnamefont {G.}~\bibnamefont
  {Casati}}, \bibinfo {author} {\bibfnamefont {B.~V.}\ \bibnamefont
  {Chirikov}}, \ and\ \bibinfo {author} {\bibfnamefont {I.}~\bibnamefont
  {Guarneri}},\ }\bibfield  {title} {\enquote {\bibinfo {title} {Energy-level
  statistics of integrable quantum systems},}\ }\href {\doibase
  10.1103/PhysRevLett.54.1350} {\bibfield  {journal} {\bibinfo  {journal}
  {Phys. Rev. Lett.}\ }\textbf {\bibinfo {volume} {54}},\ \bibinfo {pages}
  {1350--1353} (\bibinfo {year} {1985})}\BibitemShut {NoStop}%
\bibitem [{\citenamefont {Rela\~no}\ \emph {et~al.}(2004)\citenamefont
  {Rela\~no}, \citenamefont {Dukelsky}, \citenamefont {G\'omez},\ and\
  \citenamefont {Retamosa}}]{Relano2004}%
  \BibitemOpen
  \bibfield  {author} {\bibinfo {author} {\bibfnamefont {A.}~\bibnamefont
  {Rela\~no}}, \bibinfo {author} {\bibfnamefont {J.}~\bibnamefont {Dukelsky}},
  \bibinfo {author} {\bibfnamefont {J.~M.~G.}\ \bibnamefont {G\'omez}}, \ and\
  \bibinfo {author} {\bibfnamefont {J.}~\bibnamefont {Retamosa}},\ }\bibfield
  {title} {\enquote {\bibinfo {title} {Stringent numerical test of the poisson
  distribution for finite quantum integrable {H}amiltonians},}\ }\href
  {\doibase 10.1103/PhysRevE.70.026208} {\bibfield  {journal} {\bibinfo
  {journal} {Phys. Rev. E}\ }\textbf {\bibinfo {volume} {70}},\ \bibinfo
  {pages} {026208} (\bibinfo {year} {2004})}\BibitemShut {NoStop}%
\bibitem [{\citenamefont {Benet}\ \emph {et~al.}(2003)\citenamefont {Benet},
  \citenamefont {Leyvraz},\ and\ \citenamefont {Seligman}}]{Benet2003}%
  \BibitemOpen
  \bibfield  {author} {\bibinfo {author} {\bibfnamefont {L.}~\bibnamefont
  {Benet}}, \bibinfo {author} {\bibfnamefont {F.}~\bibnamefont {Leyvraz}}, \
  and\ \bibinfo {author} {\bibfnamefont {T.~H.}\ \bibnamefont {Seligman}},\
  }\bibfield  {title} {\enquote {\bibinfo {title} {Wigner-dyson statistics for
  a class of integrable models},}\ }\href {\doibase 10.1103/PhysRevE.68.045201}
  {\bibfield  {journal} {\bibinfo  {journal} {Phys. Rev. E}\ }\textbf {\bibinfo
  {volume} {68}},\ \bibinfo {pages} {045201(R)} (\bibinfo {year}
  {2003})}\BibitemShut {NoStop}%
\bibitem [{\citenamefont {Borgonovi}\ \emph {et~al.}(2016)\citenamefont
  {Borgonovi}, \citenamefont {Izrailev}, \citenamefont {Santos},\ and\
  \citenamefont {Zelevinsky}}]{Borgonovi2016}%
  \BibitemOpen
  \bibfield  {author} {\bibinfo {author} {\bibfnamefont {F.}~\bibnamefont
  {Borgonovi}}, \bibinfo {author} {\bibfnamefont {F.~M.}\ \bibnamefont
  {Izrailev}}, \bibinfo {author} {\bibfnamefont {L.~F.}\ \bibnamefont
  {Santos}}, \ and\ \bibinfo {author} {\bibfnamefont {V.~G.}\ \bibnamefont
  {Zelevinsky}},\ }\bibfield  {title} {\enquote {\bibinfo {title} {Quantum
  chaos and thermalization in isolated systems of interacting particles},}\
  }\href {\doibase 10.1016/j.physrep.2016.02.005} {\bibfield  {journal}
  {\bibinfo  {journal} {Phys. Rep.}\ }\textbf {\bibinfo {volume} {626}},\
  \bibinfo {pages} {1} (\bibinfo {year} {2016})}\BibitemShut {NoStop}%
\bibitem [{\citenamefont {D'Alessio}\ \emph {et~al.}(2016)\citenamefont
  {D'Alessio}, \citenamefont {Kafri}, \citenamefont {Polkovnikov},\ and\
  \citenamefont {Rigol}}]{Dalessio2016}%
  \BibitemOpen
  \bibfield  {author} {\bibinfo {author} {\bibfnamefont {Luca}\ \bibnamefont
  {D'Alessio}}, \bibinfo {author} {\bibfnamefont {Yariv}\ \bibnamefont
  {Kafri}}, \bibinfo {author} {\bibfnamefont {Anatoli}\ \bibnamefont
  {Polkovnikov}}, \ and\ \bibinfo {author} {\bibfnamefont {Marcos}\
  \bibnamefont {Rigol}},\ }\bibfield  {title} {\enquote {\bibinfo {title} {From
  quantum chaos and eigenstate thermalization to statistical mechanics and
  thermodynamics},}\ }\href {\doibase 10.1080/00018732.2016.1198134} {\bibfield
   {journal} {\bibinfo  {journal} {Advances in Physics}\ }\textbf {\bibinfo
  {volume} {65}},\ \bibinfo {pages} {239--362} (\bibinfo {year} {2016})},\
  \Eprint {http://arxiv.org/abs/https://doi.org/10.1080/00018732.2016.1198134}
  {https://doi.org/10.1080/00018732.2016.1198134} \BibitemShut {NoStop}%
\bibitem [{\citenamefont {Landsman}\ \emph {et~al.}(2019)\citenamefont
  {Landsman}, \citenamefont {Figgatt}, \citenamefont {Schuster}, \citenamefont
  {Linke}, \citenamefont {Yoshida}, \citenamefont {Yao},\ and\ \citenamefont
  {Monroe}}]{Landsman2019}%
  \BibitemOpen
  \bibfield  {author} {\bibinfo {author} {\bibfnamefont {K.~A.}\ \bibnamefont
  {Landsman}}, \bibinfo {author} {\bibfnamefont {C.}~\bibnamefont {Figgatt}},
  \bibinfo {author} {\bibfnamefont {T.}~\bibnamefont {Schuster}}, \bibinfo
  {author} {\bibfnamefont {N.~M.}\ \bibnamefont {Linke}}, \bibinfo {author}
  {\bibfnamefont {B.}~\bibnamefont {Yoshida}}, \bibinfo {author} {\bibfnamefont
  {N.~Y.}\ \bibnamefont {Yao}}, \ and\ \bibinfo {author} {\bibfnamefont
  {C.}~\bibnamefont {Monroe}},\ }\bibfield  {title} {\enquote {\bibinfo {title}
  {Verified quantum information scrambling},}\ }\href {\doibase
  10.1038/s41586-019-0952-6} {\bibfield  {journal} {\bibinfo  {journal}
  {Nature}\ }\textbf {\bibinfo {volume} {567}},\ \bibinfo {pages} {61--65}
  (\bibinfo {year} {2019})}\BibitemShut {NoStop}%
\bibitem [{\citenamefont {\v{S}untajs}\ \emph {et~al.}(2020)\citenamefont
  {\v{S}untajs}, \citenamefont {Bon\v{c}a}, \citenamefont {Prosen},\ and\
  \citenamefont {Vidmar}}]{Suntajs2020}%
  \BibitemOpen
  \bibfield  {author} {\bibinfo {author} {\bibfnamefont {Jan}\ \bibnamefont
  {\v{S}untajs}}, \bibinfo {author} {\bibfnamefont {Janez}\ \bibnamefont
  {Bon\v{c}a}}, \bibinfo {author} {\bibfnamefont {Toma\v{z}}\ \bibnamefont
  {Prosen}}, \ and\ \bibinfo {author} {\bibfnamefont {Lev}\ \bibnamefont
  {Vidmar}},\ }\bibfield  {title} {\enquote {\bibinfo {title} {Quantum chaos
  challenges many-body localization},}\ }\href {\doibase
  10.1103/PhysRevE.102.062144} {\bibfield  {journal} {\bibinfo  {journal}
  {Phys. Rev. E}\ }\textbf {\bibinfo {volume} {102}},\ \bibinfo {pages}
  {062144} (\bibinfo {year} {2020})}\BibitemShut {NoStop}%
\bibitem [{\citenamefont {Jensen}(2016)}]{Jensen2016}%
  \BibitemOpen
  \bibfield  {author} {\bibinfo {author} {\bibfnamefont {Kristan}\ \bibnamefont
  {Jensen}},\ }\bibfield  {title} {\enquote {\bibinfo {title} {Chaos in
  ${\mathrm{ads}}_{2}$ holography},}\ }\href {\doibase
  10.1103/PhysRevLett.117.111601} {\bibfield  {journal} {\bibinfo  {journal}
  {Phys. Rev. Lett.}\ }\textbf {\bibinfo {volume} {117}},\ \bibinfo {pages}
  {111601} (\bibinfo {year} {2016})}\BibitemShut {NoStop}%
\bibitem [{\citenamefont {Denisov}\ \emph {et~al.}(2019)\citenamefont
  {Denisov}, \citenamefont {Laptyeva}, \citenamefont {Tarnowski}, \citenamefont
  {Chru\ifmmode \acute{s}\else \'{s}\fi{}ci\ifmmode~\acute{n}\else
  \'{n}\fi{}ski},\ and\ \citenamefont {\ifmmode~\dot{Z}\else
  \.{Z}\fi{}yczkowski}}]{Denisov2019}%
  \BibitemOpen
  \bibfield  {author} {\bibinfo {author} {\bibfnamefont {Sergey}\ \bibnamefont
  {Denisov}}, \bibinfo {author} {\bibfnamefont {Tetyana}\ \bibnamefont
  {Laptyeva}}, \bibinfo {author} {\bibfnamefont {Wojciech}\ \bibnamefont
  {Tarnowski}}, \bibinfo {author} {\bibfnamefont {Dariusz}\ \bibnamefont
  {Chru\ifmmode \acute{s}\else \'{s}\fi{}ci\ifmmode~\acute{n}\else
  \'{n}\fi{}ski}}, \ and\ \bibinfo {author} {\bibfnamefont {Karol}\
  \bibnamefont {\ifmmode~\dot{Z}\else \.{Z}\fi{}yczkowski}},\ }\bibfield
  {title} {\enquote {\bibinfo {title} {Universal spectra of random {L}indblad
  operators},}\ }\href {\doibase 10.1103/PhysRevLett.123.140403} {\bibfield
  {journal} {\bibinfo  {journal} {Phys. Rev. Lett.}\ }\textbf {\bibinfo
  {volume} {123}},\ \bibinfo {pages} {140403} (\bibinfo {year}
  {2019})}\BibitemShut {NoStop}%
\bibitem [{\citenamefont {Akemann}\ \emph {et~al.}(2019)\citenamefont
  {Akemann}, \citenamefont {Kieburg}, \citenamefont {Mielke},\ and\
  \citenamefont {Prosen}}]{Akemann2019}%
  \BibitemOpen
  \bibfield  {author} {\bibinfo {author} {\bibfnamefont {Gernot}\ \bibnamefont
  {Akemann}}, \bibinfo {author} {\bibfnamefont {Mario}\ \bibnamefont
  {Kieburg}}, \bibinfo {author} {\bibfnamefont {Adam}\ \bibnamefont {Mielke}},
  \ and\ \bibinfo {author} {\bibfnamefont {Toma\v{z}}\ \bibnamefont {Prosen}},\
  }\bibfield  {title} {\enquote {\bibinfo {title} {Universal signature from
  integrability to chaos in dissipative open quantum systems},}\ }\href
  {\doibase 10.1103/PhysRevLett.123.254101} {\bibfield  {journal} {\bibinfo
  {journal} {Phys. Rev. Lett.}\ }\textbf {\bibinfo {volume} {123}},\ \bibinfo
  {pages} {254101} (\bibinfo {year} {2019})}\BibitemShut {NoStop}%
\bibitem [{\citenamefont {S\'a}\ \emph {et~al.}(2020)\citenamefont {S\'a},
  \citenamefont {Ribeiro},\ and\ \citenamefont {Prosen}}]{Sa2020}%
  \BibitemOpen
  \bibfield  {author} {\bibinfo {author} {\bibfnamefont {Lucas}\ \bibnamefont
  {S\'a}}, \bibinfo {author} {\bibfnamefont {Pedro}\ \bibnamefont {Ribeiro}}, \
  and\ \bibinfo {author} {\bibfnamefont {Toma\ifmmode\check{z}\else\v{z}\fi{}}\
  \bibnamefont {Prosen}},\ }\bibfield  {title} {\enquote {\bibinfo {title}
  {Complex spacing ratios: A signature of dissipative quantum chaos},}\ }\href
  {\doibase 10.1103/PhysRevX.10.021019} {\bibfield  {journal} {\bibinfo
  {journal} {Phys. Rev. X}\ }\textbf {\bibinfo {volume} {10}},\ \bibinfo
  {pages} {021019} (\bibinfo {year} {2020})}\BibitemShut {NoStop}%
\bibitem [{\citenamefont {Hamazaki}\ \emph {et~al.}(2020)\citenamefont
  {Hamazaki}, \citenamefont {Kawabata}, \citenamefont {Kura},\ and\
  \citenamefont {Ueda}}]{Hamazaki2020}%
  \BibitemOpen
  \bibfield  {author} {\bibinfo {author} {\bibfnamefont {Ryusuke}\ \bibnamefont
  {Hamazaki}}, \bibinfo {author} {\bibfnamefont {Kohei}\ \bibnamefont
  {Kawabata}}, \bibinfo {author} {\bibfnamefont {Naoto}\ \bibnamefont {Kura}},
  \ and\ \bibinfo {author} {\bibfnamefont {Masahito}\ \bibnamefont {Ueda}},\
  }\bibfield  {title} {\enquote {\bibinfo {title} {Universality classes of
  non-{H}ermitian random matrices},}\ }\href {\doibase
  10.1103/PhysRevResearch.2.023286} {\bibfield  {journal} {\bibinfo  {journal}
  {Phys. Rev. Res.}\ }\textbf {\bibinfo {volume} {2}},\ \bibinfo {pages}
  {023286} (\bibinfo {year} {2020})}\BibitemShut {NoStop}%
\bibitem [{\citenamefont {Li}\ \emph {et~al.}(2021)\citenamefont {Li},
  \citenamefont {Prosen},\ and\ \citenamefont {Chan}}]{Li2021}%
  \BibitemOpen
  \bibfield  {author} {\bibinfo {author} {\bibfnamefont {Jiachen}\ \bibnamefont
  {Li}}, \bibinfo {author} {\bibfnamefont
  {Toma\ifmmode\check{z}\else\v{z}\fi{}}\ \bibnamefont {Prosen}}, \ and\
  \bibinfo {author} {\bibfnamefont {Amos}\ \bibnamefont {Chan}},\ }\bibfield
  {title} {\enquote {\bibinfo {title} {Spectral statistics of non-{H}ermitian
  matrices and dissipative quantum chaos},}\ }\href {\doibase
  10.1103/PhysRevLett.127.170602} {\bibfield  {journal} {\bibinfo  {journal}
  {Phys. Rev. Lett.}\ }\textbf {\bibinfo {volume} {127}},\ \bibinfo {pages}
  {170602} (\bibinfo {year} {2021})}\BibitemShut {NoStop}%
\bibitem [{\citenamefont {Rubio-Garc{\'i}a}\ \emph {et~al.}(2022)\citenamefont
  {Rubio-Garc{\'i}a}, \citenamefont {Molina},\ and\ \citenamefont
  {Dukelsky}}]{Rubio2022}%
  \BibitemOpen
  \bibfield  {author} {\bibinfo {author} {\bibfnamefont {{\'A}lvaro}\
  \bibnamefont {Rubio-Garc{\'i}a}}, \bibinfo {author} {\bibfnamefont
  {Rafael~A.}\ \bibnamefont {Molina}}, \ and\ \bibinfo {author} {\bibfnamefont
  {Jorge}\ \bibnamefont {Dukelsky}},\ }\bibfield  {title} {\enquote {\bibinfo
  {title} {{From integrability to chaos in quantum {L}iouvillians}},}\ }\href
  {\doibase 10.21468/SciPostPhysCore.5.2.026} {\bibfield  {journal} {\bibinfo
  {journal} {SciPost Phys. Core}\ }\textbf {\bibinfo {volume} {5}},\ \bibinfo
  {pages} {026} (\bibinfo {year} {2022})}\BibitemShut {NoStop}%
\bibitem [{\citenamefont {Garc\'{\i}a-Garc\'{\i}a}\ \emph
  {et~al.}(2022)\citenamefont {Garc\'{\i}a-Garc\'{\i}a}, \citenamefont
  {S{\'a}},\ and\ \citenamefont {Verbaarschot}}]{Garcia2022}%
  \BibitemOpen
  \bibfield  {author} {\bibinfo {author} {\bibfnamefont {Antonio~M.}\
  \bibnamefont {Garc\'{\i}a-Garc\'{\i}a}}, \bibinfo {author} {\bibfnamefont
  {Lucas}\ \bibnamefont {S{\'a}}}, \ and\ \bibinfo {author} {\bibfnamefont
  {Jacobus J.~M.}\ \bibnamefont {Verbaarschot}},\ }\bibfield  {title} {\enquote
  {\bibinfo {title} {Symmetry classification and universality in
  non-{H}ermitian many-body quantum chaos by the {S}achdev-{Y}e-{K}itaev
  model},}\ }\href {\doibase 10.1103/PhysRevX.12.021040} {\bibfield  {journal}
  {\bibinfo  {journal} {Phys. Rev. X}\ }\textbf {\bibinfo {volume} {12}},\
  \bibinfo {pages} {021040} (\bibinfo {year} {2022})}\BibitemShut {NoStop}%
\bibitem [{\citenamefont {Kawabata}\ \emph {et~al.}(2023)\citenamefont
  {Kawabata}, \citenamefont {Kulkarni}, \citenamefont {Li}, \citenamefont
  {Numasawa},\ and\ \citenamefont {Ryu}}]{Kawabata2023}%
  \BibitemOpen
  \bibfield  {author} {\bibinfo {author} {\bibfnamefont {Kohei}\ \bibnamefont
  {Kawabata}}, \bibinfo {author} {\bibfnamefont {Anish}\ \bibnamefont
  {Kulkarni}}, \bibinfo {author} {\bibfnamefont {Jiachen}\ \bibnamefont {Li}},
  \bibinfo {author} {\bibfnamefont {Tokiro}\ \bibnamefont {Numasawa}}, \ and\
  \bibinfo {author} {\bibfnamefont {Shinsei}\ \bibnamefont {Ryu}},\ }\bibfield
  {title} {\enquote {\bibinfo {title} {Symmetry of open quantum systems:
  Classification of dissipative quantum chaos},}\ }\href {\doibase
  10.1103/PRXQuantum.4.030328} {\bibfield  {journal} {\bibinfo  {journal} {PRX
  Quantum}\ }\textbf {\bibinfo {volume} {4}},\ \bibinfo {pages} {030328}
  (\bibinfo {year} {2023})}\BibitemShut {NoStop}%
\bibitem [{\citenamefont {S{\'a}}\ \emph {et~al.}(2023)\citenamefont {S{\'a}},
  \citenamefont {Ribeiro},\ and\ \citenamefont {Prosen}}]{Sa2023}%
  \BibitemOpen
  \bibfield  {author} {\bibinfo {author} {\bibfnamefont {Lucas}\ \bibnamefont
  {S{\'a}}}, \bibinfo {author} {\bibfnamefont {Pedro}\ \bibnamefont {Ribeiro}},
  \ and\ \bibinfo {author} {\bibfnamefont
  {Toma\ifmmode\check{z}\else\v{z}\fi{}}\ \bibnamefont {Prosen}},\ }\bibfield
  {title} {\enquote {\bibinfo {title} {Symmetry classification of many-body
  {L}indbladians: Tenfold way and beyond},}\ }\href {\doibase
  10.1103/PhysRevX.13.031019} {\bibfield  {journal} {\bibinfo  {journal} {Phys.
  Rev. X}\ }\textbf {\bibinfo {volume} {13}},\ \bibinfo {pages} {031019}
  (\bibinfo {year} {2023})}\BibitemShut {NoStop}%
\bibitem [{\citenamefont {Garc\'{\i}a-Garc\'{\i}a}\ \emph
  {et~al.}(2023)\citenamefont {Garc\'{\i}a-Garc\'{\i}a}, \citenamefont
  {S{\'a}},\ and\ \citenamefont {Verbaarschot}}]{Garcia2023}%
  \BibitemOpen
  \bibfield  {author} {\bibinfo {author} {\bibfnamefont {Antonio~M.}\
  \bibnamefont {Garc\'{\i}a-Garc\'{\i}a}}, \bibinfo {author} {\bibfnamefont
  {Lucas}\ \bibnamefont {S{\'a}}}, \ and\ \bibinfo {author} {\bibfnamefont
  {Jacobus J.~M.}\ \bibnamefont {Verbaarschot}},\ }\bibfield  {title} {\enquote
  {\bibinfo {title} {Universality and its limits in non-{H}ermitian many-body
  quantum chaos using the {S}achdev-{Y}e-{K}itaev model},}\ }\href {\doibase
  10.1103/PhysRevD.107.066007} {\bibfield  {journal} {\bibinfo  {journal}
  {Phys. Rev. D}\ }\textbf {\bibinfo {volume} {107}},\ \bibinfo {pages}
  {066007} (\bibinfo {year} {2023})}\BibitemShut {NoStop}%
\bibitem [{\citenamefont {Ferrari}\ \emph {et~al.}(2023)\citenamefont
  {Ferrari}, \citenamefont {Gravina}, \citenamefont {Eeltink}, \citenamefont
  {Scarlino}, \citenamefont {Savona},\ and\ \citenamefont
  {Minganti}}]{Ferrari2023Arxiv}%
  \BibitemOpen
  \bibfield  {author} {\bibinfo {author} {\bibfnamefont {Filippo}\ \bibnamefont
  {Ferrari}}, \bibinfo {author} {\bibfnamefont {Luca}\ \bibnamefont {Gravina}},
  \bibinfo {author} {\bibfnamefont {Debbie}\ \bibnamefont {Eeltink}}, \bibinfo
  {author} {\bibfnamefont {Pasquale}\ \bibnamefont {Scarlino}}, \bibinfo
  {author} {\bibfnamefont {Vincenzo}\ \bibnamefont {Savona}}, \ and\ \bibinfo
  {author} {\bibfnamefont {Fabrizio}\ \bibnamefont {Minganti}},\ }\href
  {https://arxiv.org/abs/2305.15479} {\enquote {\bibinfo {title} {Steady-state
  quantum chaos in open quantum systems},}\ } (\bibinfo {year} {2023}),\
  \Eprint {http://arxiv.org/abs/2305.15479} {arXiv:2305.15479 [quant-ph]}
  \BibitemShut {NoStop}%
\bibitem [{\citenamefont {Grobe}\ \emph {et~al.}(1988)\citenamefont {Grobe},
  \citenamefont {Haake},\ and\ \citenamefont {Sommers}}]{Grobe1988}%
  \BibitemOpen
  \bibfield  {author} {\bibinfo {author} {\bibfnamefont {Rainer}\ \bibnamefont
  {Grobe}}, \bibinfo {author} {\bibfnamefont {Fritz}\ \bibnamefont {Haake}}, \
  and\ \bibinfo {author} {\bibfnamefont {Hans-J\"urgen}\ \bibnamefont
  {Sommers}},\ }\bibfield  {title} {\enquote {\bibinfo {title} {Quantum
  distinction of regular and chaotic dissipative motion},}\ }\href {\doibase
  10.1103/PhysRevLett.61.1899} {\bibfield  {journal} {\bibinfo  {journal}
  {Phys. Rev. Lett.}\ }\textbf {\bibinfo {volume} {61}},\ \bibinfo {pages}
  {1899--1902} (\bibinfo {year} {1988})}\BibitemShut {NoStop}%
\bibitem [{\citenamefont {Guckenheimer}\ and\ \citenamefont
  {Holmes}(1983)}]{Guckenheimer1983book}%
  \BibitemOpen
  \bibfield  {author} {\bibinfo {author} {\bibfnamefont {John}\ \bibnamefont
  {Guckenheimer}}\ and\ \bibinfo {author} {\bibfnamefont {Philip}\ \bibnamefont
  {Holmes}},\ }\href@noop {} {\emph {\bibinfo {title} {Nonlinear Oscillations,
  Dynamical Systems, and Bifurcations of Vector Fields}}}\ (\bibinfo
  {publisher} {Springer},\ \bibinfo {address} {New York},\ \bibinfo {year}
  {1983})\BibitemShut {NoStop}%
\bibitem [{\citenamefont {Shivamoggi}(2014)}]{Shivamoggi2014book}%
  \BibitemOpen
  \bibfield  {author} {\bibinfo {author} {\bibfnamefont {Bhimsen~K.}\
  \bibnamefont {Shivamoggi}},\ }\href@noop {} {\emph {\bibinfo {title}
  {Nonlinear Dynamics and Chaotic Phenomena: An Introduction}}}\ (\bibinfo
  {publisher} {Springer},\ \bibinfo {address} {Dordrecht},\ \bibinfo {year}
  {2014})\BibitemShut {NoStop}%
\bibitem [{\citenamefont {Smyrlis}\ and\ \citenamefont
  {Papageorgiou}(1991)}]{Smyrlis1991}%
  \BibitemOpen
  \bibfield  {author} {\bibinfo {author} {\bibfnamefont {Y~S}\ \bibnamefont
  {Smyrlis}}\ and\ \bibinfo {author} {\bibfnamefont {D~T}\ \bibnamefont
  {Papageorgiou}},\ }\bibfield  {title} {\enquote {\bibinfo {title} {Predicting
  chaos for infinite dimensional dynamical systems: the
  {K}uramoto-{S}ivashinsky equation, a case study},}\ }\href {\doibase
  10.1073/pnas.88.24.11129} {\bibfield  {journal} {\bibinfo  {journal} {Proc.
  Nat. Acad. Sci.}\ }\textbf {\bibinfo {volume} {88}},\ \bibinfo {pages}
  {11129--11132} (\bibinfo {year} {1991})}\BibitemShut {NoStop}%
\bibitem [{\citenamefont {Papageorgiou}\ and\ \citenamefont
  {Smyrlis}(1991)}]{Papageorgiou1991}%
  \BibitemOpen
  \bibfield  {author} {\bibinfo {author} {\bibfnamefont {Demetrios~T.}\
  \bibnamefont {Papageorgiou}}\ and\ \bibinfo {author} {\bibfnamefont
  {Yiorgos~S.}\ \bibnamefont {Smyrlis}},\ }\bibfield  {title} {\enquote
  {\bibinfo {title} {The route to chaos for the {K}uramoto-{S}ivashinsky
  equation},}\ }\href {\doibase 10.1007/BF00271514} {\bibfield  {journal}
  {\bibinfo  {journal} {Theor. Comp. Fluid Dynamics}\ }\textbf {\bibinfo
  {volume} {3}},\ \bibinfo {pages} {15--42} (\bibinfo {year}
  {1991})}\BibitemShut {NoStop}%
\bibitem [{\citenamefont {Grobe}\ and\ \citenamefont
  {Haake}(1989)}]{Grobe1989}%
  \BibitemOpen
  \bibfield  {author} {\bibinfo {author} {\bibfnamefont {Rainer}\ \bibnamefont
  {Grobe}}\ and\ \bibinfo {author} {\bibfnamefont {Fritz}\ \bibnamefont
  {Haake}},\ }\bibfield  {title} {\enquote {\bibinfo {title} {Universality of
  cubic-level repulsion for dissipative quantum chaos},}\ }\href {\doibase
  10.1103/PhysRevLett.62.2893} {\bibfield  {journal} {\bibinfo  {journal}
  {Phys. Rev. Lett.}\ }\textbf {\bibinfo {volume} {62}},\ \bibinfo {pages}
  {2893--2896} (\bibinfo {year} {1989})}\BibitemShut {NoStop}%
\bibitem [{\citenamefont {Breuer}\ and\ \citenamefont
  {Petruccione}(2002)}]{BreuerBook}%
  \BibitemOpen
  \bibfield  {author} {\bibinfo {author} {\bibfnamefont {H.-P.}\ \bibnamefont
  {Breuer}}\ and\ \bibinfo {author} {\bibfnamefont {F.}~\bibnamefont
  {Petruccione}},\ }\href@noop {} {\emph {\bibinfo {title} {The Theory of Open
  Quantum Systems}}}\ (\bibinfo  {publisher} {Oxford University},\ \bibinfo
  {address} {New York, NY},\ \bibinfo {year} {2002})\BibitemShut {NoStop}%
\bibitem [{\citenamefont {Carmichael}(1993)}]{CarmichaelBook1993}%
  \BibitemOpen
  \bibfield  {author} {\bibinfo {author} {\bibfnamefont {Howard}\ \bibnamefont
  {Carmichael}},\ }\href@noop {} {\emph {\bibinfo {title} {An Open Systems
  Approach to Quantum Optics. {L}ectures Presented at the Universit{\'e} Libre
  de Bruxelles, October 28 to November 4, 1991}}}\ (\bibinfo  {publisher}
  {Springer-Verlag},\ \bibinfo {address} {Berlin},\ \bibinfo {year}
  {1993})\BibitemShut {NoStop}%
\bibitem [{\citenamefont {Carmichael}(2002)}]{CarmichaelBook2002}%
  \BibitemOpen
  \bibfield  {author} {\bibinfo {author} {\bibfnamefont {H.~J.}\ \bibnamefont
  {Carmichael}},\ }\href@noop {} {\emph {\bibinfo {title} {Statistical Methods
  in Quantum Optics 1: {M}aster Equations and Fokker-Planck Equations}}}\
  (\bibinfo  {publisher} {Springer-Verlag},\ \bibinfo {address} {Berlin},\
  \bibinfo {year} {2002})\BibitemShut {NoStop}%
\bibitem [{\citenamefont {Jaiswal}\ \emph {et~al.}(2019)\citenamefont
  {Jaiswal}, \citenamefont {Pandey},\ and\ \citenamefont
  {Prakash}}]{Jaiswal2019}%
  \BibitemOpen
  \bibfield  {author} {\bibinfo {author} {\bibfnamefont {Ambuja~Bhushan}\
  \bibnamefont {Jaiswal}}, \bibinfo {author} {\bibfnamefont {Akhilesh}\
  \bibnamefont {Pandey}}, \ and\ \bibinfo {author} {\bibfnamefont {Ravi}\
  \bibnamefont {Prakash}},\ }\bibfield  {title} {\enquote {\bibinfo {title}
  {Universality classes of quantum chaotic dissipative systems},}\ }\href
  {\doibase 10.1209/0295-5075/127/30004} {\bibfield  {journal} {\bibinfo
  {journal} {Europhys. Lett.}\ }\textbf {\bibinfo {volume} {127}},\ \bibinfo
  {pages} {30004} (\bibinfo {year} {2019})}\BibitemShut {NoStop}%
\bibitem [{\citenamefont {Hamazaki}\ \emph {et~al.}(2019)\citenamefont
  {Hamazaki}, \citenamefont {Kawabata},\ and\ \citenamefont
  {Ueda}}]{Hamazaki2019}%
  \BibitemOpen
  \bibfield  {author} {\bibinfo {author} {\bibfnamefont {Ryusuke}\ \bibnamefont
  {Hamazaki}}, \bibinfo {author} {\bibfnamefont {Kohei}\ \bibnamefont
  {Kawabata}}, \ and\ \bibinfo {author} {\bibfnamefont {Masahito}\ \bibnamefont
  {Ueda}},\ }\bibfield  {title} {\enquote {\bibinfo {title} {Non-{H}ermitian
  many-body localization},}\ }\href {\doibase 10.1103/PhysRevLett.123.090603}
  {\bibfield  {journal} {\bibinfo  {journal} {Phys. Rev. Lett.}\ }\textbf
  {\bibinfo {volume} {123}},\ \bibinfo {pages} {090603} (\bibinfo {year}
  {2019})}\BibitemShut {NoStop}%
\bibitem [{\citenamefont {Jaako}\ \emph {et~al.}(2016)\citenamefont {Jaako},
  \citenamefont {Xiang}, \citenamefont {Garcia-Ripoll},\ and\ \citenamefont
  {Rabl}}]{Jaako2016}%
  \BibitemOpen
  \bibfield  {author} {\bibinfo {author} {\bibfnamefont {Tuomas}\ \bibnamefont
  {Jaako}}, \bibinfo {author} {\bibfnamefont {Ze-Liang}\ \bibnamefont {Xiang}},
  \bibinfo {author} {\bibfnamefont {Juan~Jos\'e}\ \bibnamefont
  {Garcia-Ripoll}}, \ and\ \bibinfo {author} {\bibfnamefont {Peter}\
  \bibnamefont {Rabl}},\ }\bibfield  {title} {\enquote {\bibinfo {title}
  {Ultrastrong-coupling phenomena beyond the {D}icke model},}\ }\href {\doibase
  10.1103/PhysRevA.94.033850} {\bibfield  {journal} {\bibinfo  {journal} {Phys.
  Rev. A}\ }\textbf {\bibinfo {volume} {94}},\ \bibinfo {pages} {033850}
  (\bibinfo {year} {2016})}\BibitemShut {NoStop}%
\bibitem [{\citenamefont {Baden}\ \emph {et~al.}(2014)\citenamefont {Baden},
  \citenamefont {Arnold}, \citenamefont {Grimsmo}, \citenamefont {Parkins},\
  and\ \citenamefont {Barrett}}]{Baden2014}%
  \BibitemOpen
  \bibfield  {author} {\bibinfo {author} {\bibfnamefont {Markus~P.}\
  \bibnamefont {Baden}}, \bibinfo {author} {\bibfnamefont {Kyle~J.}\
  \bibnamefont {Arnold}}, \bibinfo {author} {\bibfnamefont {Arne~L.}\
  \bibnamefont {Grimsmo}}, \bibinfo {author} {\bibfnamefont {Scott}\
  \bibnamefont {Parkins}}, \ and\ \bibinfo {author} {\bibfnamefont {Murray~D.}\
  \bibnamefont {Barrett}},\ }\bibfield  {title} {\enquote {\bibinfo {title}
  {Realization of the {D}icke model using cavity-assisted {R}aman
  transitions},}\ }\href {\doibase 10.1103/PhysRevLett.113.020408} {\bibfield
  {journal} {\bibinfo  {journal} {Phys. Rev. Lett.}\ }\textbf {\bibinfo
  {volume} {113}},\ \bibinfo {pages} {020408} (\bibinfo {year}
  {2014})}\BibitemShut {NoStop}%
\bibitem [{\citenamefont {Klinder}\ \emph {et~al.}(2015)\citenamefont
  {Klinder}, \citenamefont {Keßler}, \citenamefont {Wolke}, \citenamefont
  {Mathey},\ and\ \citenamefont {Hemmerich}}]{Klinder2015PNAS}%
  \BibitemOpen
  \bibfield  {author} {\bibinfo {author} {\bibfnamefont {Jens}\ \bibnamefont
  {Klinder}}, \bibinfo {author} {\bibfnamefont {Hans}\ \bibnamefont {Keßler}},
  \bibinfo {author} {\bibfnamefont {Matthias}\ \bibnamefont {Wolke}}, \bibinfo
  {author} {\bibfnamefont {Ludwig}\ \bibnamefont {Mathey}}, \ and\ \bibinfo
  {author} {\bibfnamefont {Andreas}\ \bibnamefont {Hemmerich}},\ }\bibfield
  {title} {\enquote {\bibinfo {title} {Dynamical phase transition in the open
  {D}icke model},}\ }\href {\doibase 10.1073/pnas.1417132112} {\bibfield
  {journal} {\bibinfo  {journal} {Proc. Nat. Ac. Sci.}\ }\textbf {\bibinfo
  {volume} {112}},\ \bibinfo {pages} {3290--3295} (\bibinfo {year}
  {2015})}\BibitemShut {NoStop}%
\bibitem [{\citenamefont {Zhiqiang}\ \emph {et~al.}(2017)\citenamefont
  {Zhiqiang}, \citenamefont {Lee}, \citenamefont {Kumar}, \citenamefont
  {Arnold}, \citenamefont {Masson}, \citenamefont {Parkins},\ and\
  \citenamefont {Barrett}}]{Zhiqiang2017}%
  \BibitemOpen
  \bibfield  {author} {\bibinfo {author} {\bibfnamefont {Zhang}\ \bibnamefont
  {Zhiqiang}}, \bibinfo {author} {\bibfnamefont {Chern~Hui}\ \bibnamefont
  {Lee}}, \bibinfo {author} {\bibfnamefont {Ravi}\ \bibnamefont {Kumar}},
  \bibinfo {author} {\bibfnamefont {K.~J.}\ \bibnamefont {Arnold}}, \bibinfo
  {author} {\bibfnamefont {Stuart~J.}\ \bibnamefont {Masson}}, \bibinfo
  {author} {\bibfnamefont {A.~S.}\ \bibnamefont {Parkins}}, \ and\ \bibinfo
  {author} {\bibfnamefont {M.~D.}\ \bibnamefont {Barrett}},\ }\bibfield
  {title} {\enquote {\bibinfo {title} {Nonequilibrium phase transition in a
  spin-1 {D}icke model},}\ }\href {\doibase 10.1364/OPTICA.4.000424} {\bibfield
   {journal} {\bibinfo  {journal} {Optica}\ }\textbf {\bibinfo {volume} {4}},\
  \bibinfo {pages} {424--429} (\bibinfo {year} {2017})}\BibitemShut {NoStop}%
\bibitem [{\citenamefont {Zhang}\ \emph {et~al.}(2018)\citenamefont {Zhang},
  \citenamefont {Lee}, \citenamefont {Kumar}, \citenamefont {Arnold},
  \citenamefont {Masson}, \citenamefont {Grimsmo}, \citenamefont {Parkins},\
  and\ \citenamefont {Barrett}}]{Zhang2018}%
  \BibitemOpen
  \bibfield  {author} {\bibinfo {author} {\bibfnamefont {Zhiqiang}\
  \bibnamefont {Zhang}}, \bibinfo {author} {\bibfnamefont {Chern~Hui}\
  \bibnamefont {Lee}}, \bibinfo {author} {\bibfnamefont {Ravi}\ \bibnamefont
  {Kumar}}, \bibinfo {author} {\bibfnamefont {K.~J.}\ \bibnamefont {Arnold}},
  \bibinfo {author} {\bibfnamefont {Stuart~J.}\ \bibnamefont {Masson}},
  \bibinfo {author} {\bibfnamefont {A.~L.}\ \bibnamefont {Grimsmo}}, \bibinfo
  {author} {\bibfnamefont {A.~S.}\ \bibnamefont {Parkins}}, \ and\ \bibinfo
  {author} {\bibfnamefont {M.~D.}\ \bibnamefont {Barrett}},\ }\bibfield
  {title} {\enquote {\bibinfo {title} {Dicke-model simulation via
  cavity-assisted {R}aman transitions},}\ }\href {\doibase
  10.1103/PhysRevA.97.043858} {\bibfield  {journal} {\bibinfo  {journal} {Phys.
  Rev. A}\ }\textbf {\bibinfo {volume} {97}},\ \bibinfo {pages} {043858}
  (\bibinfo {year} {2018})}\BibitemShut {NoStop}%
\bibitem [{\citenamefont {Cohn}\ \emph {et~al.}(2018)\citenamefont {Cohn},
  \citenamefont {Safavi-Naini}, \citenamefont {Lewis-Swan}, \citenamefont
  {Bohnet}, \citenamefont {G\"arttner}, \citenamefont {Gilmore}, \citenamefont
  {Jordan}, \citenamefont {Rey}, \citenamefont {Bollinger},\ and\ \citenamefont
  {Freericks}}]{Cohn2018}%
  \BibitemOpen
  \bibfield  {author} {\bibinfo {author} {\bibfnamefont {J}~\bibnamefont
  {Cohn}}, \bibinfo {author} {\bibfnamefont {A}~\bibnamefont {Safavi-Naini}},
  \bibinfo {author} {\bibfnamefont {R~J}\ \bibnamefont {Lewis-Swan}}, \bibinfo
  {author} {\bibfnamefont {J~G}\ \bibnamefont {Bohnet}}, \bibinfo {author}
  {\bibfnamefont {M}~\bibnamefont {G\"arttner}}, \bibinfo {author}
  {\bibfnamefont {K~A}\ \bibnamefont {Gilmore}}, \bibinfo {author}
  {\bibfnamefont {J~E}\ \bibnamefont {Jordan}}, \bibinfo {author}
  {\bibfnamefont {A~M}\ \bibnamefont {Rey}}, \bibinfo {author} {\bibfnamefont
  {J~J}\ \bibnamefont {Bollinger}}, \ and\ \bibinfo {author} {\bibfnamefont
  {J~K}\ \bibnamefont {Freericks}},\ }\bibfield  {title} {\enquote {\bibinfo
  {title} {Bang-bang shortcut to adiabaticity in the {D}icke model as realized
  in a penning trap experiment},}\ }\href
  {http://stacks.iop.org/1367-2630/20/i=5/a=055013} {\bibfield  {journal}
  {\bibinfo  {journal} {New J. Phys.}\ }\textbf {\bibinfo {volume} {20}},\
  \bibinfo {pages} {055013} (\bibinfo {year} {2018})}\BibitemShut {NoStop}%
\bibitem [{\citenamefont {Safavi-Naini}\ \emph {et~al.}(2018)\citenamefont
  {Safavi-Naini}, \citenamefont {Lewis-Swan}, \citenamefont {Bohnet},
  \citenamefont {G\"arttner}, \citenamefont {Gilmore}, \citenamefont {Jordan},
  \citenamefont {Cohn}, \citenamefont {Freericks}, \citenamefont {Rey},\ and\
  \citenamefont {Bollinger}}]{Safavi2018}%
  \BibitemOpen
  \bibfield  {author} {\bibinfo {author} {\bibfnamefont {A.}~\bibnamefont
  {Safavi-Naini}}, \bibinfo {author} {\bibfnamefont {R.~J.}\ \bibnamefont
  {Lewis-Swan}}, \bibinfo {author} {\bibfnamefont {J.~G.}\ \bibnamefont
  {Bohnet}}, \bibinfo {author} {\bibfnamefont {M.}~\bibnamefont {G\"arttner}},
  \bibinfo {author} {\bibfnamefont {K.~A.}\ \bibnamefont {Gilmore}}, \bibinfo
  {author} {\bibfnamefont {J.~E.}\ \bibnamefont {Jordan}}, \bibinfo {author}
  {\bibfnamefont {J.}~\bibnamefont {Cohn}}, \bibinfo {author} {\bibfnamefont
  {J.~K.}\ \bibnamefont {Freericks}}, \bibinfo {author} {\bibfnamefont {A.~M.}\
  \bibnamefont {Rey}}, \ and\ \bibinfo {author} {\bibfnamefont {J.~J.}\
  \bibnamefont {Bollinger}},\ }\bibfield  {title} {\enquote {\bibinfo {title}
  {Verification of a many-ion simulator of the {D}icke model through slow
  quenches across a phase transition},}\ }\href {\doibase
  10.1103/PhysRevLett.121.040503} {\bibfield  {journal} {\bibinfo  {journal}
  {Phys. Rev. Lett.}\ }\textbf {\bibinfo {volume} {121}},\ \bibinfo {pages}
  {040503} (\bibinfo {year} {2018})}\BibitemShut {NoStop}%
\bibitem [{\citenamefont {Dicke}(1954)}]{Dicke1954}%
  \BibitemOpen
  \bibfield  {author} {\bibinfo {author} {\bibfnamefont {R.~H.}\ \bibnamefont
  {Dicke}},\ }\bibfield  {title} {\enquote {\bibinfo {title} {Coherence in
  spontaneous radiation processes},}\ }\href {\doibase 10.1103/PhysRev.93.99}
  {\bibfield  {journal} {\bibinfo  {journal} {Phys. Rev.}\ }\textbf {\bibinfo
  {volume} {93}},\ \bibinfo {pages} {99} (\bibinfo {year} {1954})}\BibitemShut
  {NoStop}%
\bibitem [{\citenamefont {Garraway}(2011)}]{Garraway2011}%
  \BibitemOpen
  \bibfield  {author} {\bibinfo {author} {\bibfnamefont {Barry~M.}\
  \bibnamefont {Garraway}},\ }\bibfield  {title} {\enquote {\bibinfo {title}
  {The {D}icke model in quantum optics: {D}icke model revisited},}\ }\href
  {\doibase 10.1098/rsta.2010.0333} {\bibfield  {journal} {\bibinfo  {journal}
  {Philos. Trans. Royal Soc. A}\ }\textbf {\bibinfo {volume} {369}},\ \bibinfo
  {pages} {1137} (\bibinfo {year} {2011})}\BibitemShut {NoStop}%
\bibitem [{\citenamefont {Kirton}\ \emph {et~al.}(2019)\citenamefont {Kirton},
  \citenamefont {Roses}, \citenamefont {Keeling},\ and\ \citenamefont
  {Dalla~Torre}}]{Kirton2019}%
  \BibitemOpen
  \bibfield  {author} {\bibinfo {author} {\bibfnamefont {Peter}\ \bibnamefont
  {Kirton}}, \bibinfo {author} {\bibfnamefont {Mor~M.}\ \bibnamefont {Roses}},
  \bibinfo {author} {\bibfnamefont {Jonathan}\ \bibnamefont {Keeling}}, \ and\
  \bibinfo {author} {\bibfnamefont {Emanuele~G.}\ \bibnamefont {Dalla~Torre}},\
  }\bibfield  {title} {\enquote {\bibinfo {title} {Introduction to the {D}icke
  model: From equilibrium to nonequilibrium, and vice versa},}\ }\href
  {\doibase https://doi.org/10.1002/qute.201800043} {\bibfield  {journal}
  {\bibinfo  {journal} {Adv. Quantum Technol.}\ }\textbf {\bibinfo {volume}
  {2}},\ \bibinfo {pages} {1800043} (\bibinfo {year} {2019})}\BibitemShut
  {NoStop}%
\bibitem [{\citenamefont {Villase{\~n}or}\ \emph {et~al.}(2024)\citenamefont
  {Villase{\~n}or}, \citenamefont {Pilatowsky-Cameo}, \citenamefont
  {Ch{\'a}vez-Carlos}, \citenamefont {Bastarrachea-Magnani}, \citenamefont
  {Lerma-Hern{\'a}ndez}, \citenamefont {Santos},\ and\ \citenamefont
  {Hirsch}}]{Villasenor2024ARXIV}%
  \BibitemOpen
  \bibfield  {author} {\bibinfo {author} {\bibfnamefont {David}\ \bibnamefont
  {Villase{\~n}or}}, \bibinfo {author} {\bibfnamefont {Sa{\'u}l}\ \bibnamefont
  {Pilatowsky-Cameo}}, \bibinfo {author} {\bibfnamefont {Jorge}\ \bibnamefont
  {Ch{\'a}vez-Carlos}}, \bibinfo {author} {\bibfnamefont {Miguel~A.}\
  \bibnamefont {Bastarrachea-Magnani}}, \bibinfo {author} {\bibfnamefont
  {Sergio}\ \bibnamefont {Lerma-Hern{\'a}ndez}}, \bibinfo {author}
  {\bibfnamefont {Lea~F.}\ \bibnamefont {Santos}}, \ and\ \bibinfo {author}
  {\bibfnamefont {Jorge~G.}\ \bibnamefont {Hirsch}},\ }\href@noop {} {\enquote
  {\bibinfo {title} {Classical and quantum properties of the spin-boson {D}icke
  model: Chaos, localization, and scarring},}\ } (\bibinfo {year} {2024}),\
  \Eprint {http://arxiv.org/abs/2405.20381} {arXiv:2405.20381 [quant-ph]}
  \BibitemShut {NoStop}%
\bibitem [{\citenamefont {Villase\~nor}\ and\ \citenamefont
  {Barberis-Blostein}(2024)}]{Villasenor2024}%
  \BibitemOpen
  \bibfield  {author} {\bibinfo {author} {\bibfnamefont {David}\ \bibnamefont
  {Villase\~nor}}\ and\ \bibinfo {author} {\bibfnamefont {Pablo}\ \bibnamefont
  {Barberis-Blostein}},\ }\bibfield  {title} {\enquote {\bibinfo {title}
  {Analysis of chaos and regularity in the open {D}icke model},}\ }\href
  {\doibase 10.1103/PhysRevE.109.014206} {\bibfield  {journal} {\bibinfo
  {journal} {Phys. Rev. E}\ }\textbf {\bibinfo {volume} {109}},\ \bibinfo
  {pages} {014206} (\bibinfo {year} {2024})}\BibitemShut {NoStop}%
\bibitem [{\citenamefont {Stitely}\ \emph {et~al.}(2020)\citenamefont
  {Stitely}, \citenamefont {Giraldo}, \citenamefont {Krauskopf},\ and\
  \citenamefont {Parkins}}]{Stitely2020}%
  \BibitemOpen
  \bibfield  {author} {\bibinfo {author} {\bibfnamefont {Kevin~C.}\
  \bibnamefont {Stitely}}, \bibinfo {author} {\bibfnamefont {Andrus}\
  \bibnamefont {Giraldo}}, \bibinfo {author} {\bibfnamefont {Bernd}\
  \bibnamefont {Krauskopf}}, \ and\ \bibinfo {author} {\bibfnamefont {Scott}\
  \bibnamefont {Parkins}},\ }\bibfield  {title} {\enquote {\bibinfo {title}
  {Nonlinear semiclassical dynamics of the unbalanced, open {D}icke model},}\
  }\href {\doibase 10.1103/PhysRevResearch.2.033131} {\bibfield  {journal}
  {\bibinfo  {journal} {Phys. Rev. Res.}\ }\textbf {\bibinfo {volume} {2}},\
  \bibinfo {pages} {033131} (\bibinfo {year} {2020})}\BibitemShut {NoStop}%
\bibitem [{\citenamefont {Hioe}(1973)}]{Hioe1973}%
  \BibitemOpen
  \bibfield  {author} {\bibinfo {author} {\bibfnamefont {F.~T.}\ \bibnamefont
  {Hioe}},\ }\bibfield  {title} {\enquote {\bibinfo {title} {Phase transitions
  in some generalized {D}icke models of superradiance},}\ }\href {\doibase
  10.1103/PhysRevA.8.1440} {\bibfield  {journal} {\bibinfo  {journal} {Phys.
  Rev. A}\ }\textbf {\bibinfo {volume} {8}},\ \bibinfo {pages} {1440--1445}
  (\bibinfo {year} {1973})}\BibitemShut {NoStop}%
\bibitem [{\citenamefont {Hepp}\ and\ \citenamefont
  {Lieb}(1973{\natexlab{a}})}]{Hepp1973a}%
  \BibitemOpen
  \bibfield  {author} {\bibinfo {author} {\bibfnamefont {Klaus}\ \bibnamefont
  {Hepp}}\ and\ \bibinfo {author} {\bibfnamefont {Elliott~H}\ \bibnamefont
  {Lieb}},\ }\bibfield  {title} {\enquote {\bibinfo {title} {On the
  superradiant phase transition for molecules in a quantized radiation field:
  The {D}icke maser model},}\ }\href {\doibase
  https://doi.org/10.1016/0003-4916(73)90039-0} {\bibfield  {journal} {\bibinfo
   {journal} {Ann. Phys. (N.Y.)}\ }\textbf {\bibinfo {volume} {76}},\ \bibinfo
  {pages} {360 -- 404} (\bibinfo {year} {1973}{\natexlab{a}})}\BibitemShut
  {NoStop}%
\bibitem [{\citenamefont {Hepp}\ and\ \citenamefont
  {Lieb}(1973{\natexlab{b}})}]{Hepp1973b}%
  \BibitemOpen
  \bibfield  {author} {\bibinfo {author} {\bibfnamefont {Klaus}\ \bibnamefont
  {Hepp}}\ and\ \bibinfo {author} {\bibfnamefont {Elliott~H.}\ \bibnamefont
  {Lieb}},\ }\bibfield  {title} {\enquote {\bibinfo {title} {Equilibrium
  statistical mechanics of matter interacting with the quantized radiation
  field},}\ }\href {\doibase 10.1103/PhysRevA.8.2517} {\bibfield  {journal}
  {\bibinfo  {journal} {Phys. Rev. A}\ }\textbf {\bibinfo {volume} {8}},\
  \bibinfo {pages} {2517--2525} (\bibinfo {year}
  {1973}{\natexlab{b}})}\BibitemShut {NoStop}%
\bibitem [{\citenamefont {Carmichael}\ \emph {et~al.}(1973)\citenamefont
  {Carmichael}, \citenamefont {Gardiner},\ and\ \citenamefont
  {Walls}}]{Carmichael1973}%
  \BibitemOpen
  \bibfield  {author} {\bibinfo {author} {\bibfnamefont {H.J.}\ \bibnamefont
  {Carmichael}}, \bibinfo {author} {\bibfnamefont {C.W.}\ \bibnamefont
  {Gardiner}}, \ and\ \bibinfo {author} {\bibfnamefont {D.F.}\ \bibnamefont
  {Walls}},\ }\bibfield  {title} {\enquote {\bibinfo {title} {Higher order
  corrections to the {D}icke superradiant phase transition},}\ }\href {\doibase
  https://doi.org/10.1016/0375-9601(73)90679-8} {\bibfield  {journal} {\bibinfo
   {journal} {Phys. Lett. A}\ }\textbf {\bibinfo {volume} {46}},\ \bibinfo
  {pages} {47 -- 48} (\bibinfo {year} {1973})}\BibitemShut {NoStop}%
\bibitem [{\citenamefont {Wang}\ and\ \citenamefont {Hioe}(1973)}]{Wang1973}%
  \BibitemOpen
  \bibfield  {author} {\bibinfo {author} {\bibfnamefont {Y.~K.}\ \bibnamefont
  {Wang}}\ and\ \bibinfo {author} {\bibfnamefont {F.~T.}\ \bibnamefont
  {Hioe}},\ }\bibfield  {title} {\enquote {\bibinfo {title} {Phase transition
  in the {D}icke model of superradiance},}\ }\href {\doibase
  10.1103/PhysRevA.7.831} {\bibfield  {journal} {\bibinfo  {journal} {Phys.
  Rev. A}\ }\textbf {\bibinfo {volume} {7}},\ \bibinfo {pages} {831--836}
  (\bibinfo {year} {1973})}\BibitemShut {NoStop}%
\bibitem [{\citenamefont {Bu\v{c}a}\ and\ \citenamefont
  {Prosen}(2012)}]{Buca2012}%
  \BibitemOpen
  \bibfield  {author} {\bibinfo {author} {\bibfnamefont {Berislav}\
  \bibnamefont {Bu\v{c}a}}\ and\ \bibinfo {author} {\bibfnamefont {Toma\v{z}}\
  \bibnamefont {Prosen}},\ }\bibfield  {title} {\enquote {\bibinfo {title} {A
  note on symmetry reductions of the {L}indblad equation: transport in
  constrained open spin chains},}\ }\href {\doibase
  10.1088/1367-2630/14/7/073007} {\bibfield  {journal} {\bibinfo  {journal}
  {New J. Phys.}\ }\textbf {\bibinfo {volume} {14}},\ \bibinfo {pages} {073007}
  (\bibinfo {year} {2012})}\BibitemShut {NoStop}%
\bibitem [{\citenamefont {Albert}\ and\ \citenamefont
  {Jiang}(2014)}]{Albert2014}%
  \BibitemOpen
  \bibfield  {author} {\bibinfo {author} {\bibfnamefont {Victor~V.}\
  \bibnamefont {Albert}}\ and\ \bibinfo {author} {\bibfnamefont {Liang}\
  \bibnamefont {Jiang}},\ }\bibfield  {title} {\enquote {\bibinfo {title}
  {Symmetries and conserved quantities in {L}indblad master equations},}\
  }\href {\doibase 10.1103/PhysRevA.89.022118} {\bibfield  {journal} {\bibinfo
  {journal} {Phys. Rev. A}\ }\textbf {\bibinfo {volume} {89}},\ \bibinfo
  {pages} {022118} (\bibinfo {year} {2014})}\BibitemShut {NoStop}%
\bibitem [{\citenamefont {Lieu}\ \emph {et~al.}(2020)\citenamefont {Lieu},
  \citenamefont {Belyansky}, \citenamefont {Young}, \citenamefont {Lundgren},
  \citenamefont {Albert},\ and\ \citenamefont {Gorshkov}}]{Lieu2020}%
  \BibitemOpen
  \bibfield  {author} {\bibinfo {author} {\bibfnamefont {Simon}\ \bibnamefont
  {Lieu}}, \bibinfo {author} {\bibfnamefont {Ron}\ \bibnamefont {Belyansky}},
  \bibinfo {author} {\bibfnamefont {Jeremy~T.}\ \bibnamefont {Young}}, \bibinfo
  {author} {\bibfnamefont {Rex}\ \bibnamefont {Lundgren}}, \bibinfo {author}
  {\bibfnamefont {Victor~V.}\ \bibnamefont {Albert}}, \ and\ \bibinfo {author}
  {\bibfnamefont {Alexey~V.}\ \bibnamefont {Gorshkov}},\ }\bibfield  {title}
  {\enquote {\bibinfo {title} {Symmetry breaking and error correction in open
  quantum systems},}\ }\href {\doibase 10.1103/PhysRevLett.125.240405}
  {\bibfield  {journal} {\bibinfo  {journal} {Phys. Rev. Lett.}\ }\textbf
  {\bibinfo {volume} {125}},\ \bibinfo {pages} {240405} (\bibinfo {year}
  {2020})}\BibitemShut {NoStop}%
\bibitem [{\citenamefont {Dimer}\ \emph {et~al.}(2007)\citenamefont {Dimer},
  \citenamefont {Estienne}, \citenamefont {Parkins},\ and\ \citenamefont
  {Carmichael}}]{Dimer2007}%
  \BibitemOpen
  \bibfield  {author} {\bibinfo {author} {\bibfnamefont {F.}~\bibnamefont
  {Dimer}}, \bibinfo {author} {\bibfnamefont {B.}~\bibnamefont {Estienne}},
  \bibinfo {author} {\bibfnamefont {A.~S.}\ \bibnamefont {Parkins}}, \ and\
  \bibinfo {author} {\bibfnamefont {H.~J.}\ \bibnamefont {Carmichael}},\
  }\bibfield  {title} {\enquote {\bibinfo {title} {Proposed realization of the
  {D}icke-model quantum phase transition in an optical cavity {QED} system},}\
  }\href {\doibase 10.1103/PhysRevA.75.013804} {\bibfield  {journal} {\bibinfo
  {journal} {Phys. Rev. A}\ }\textbf {\bibinfo {volume} {75}},\ \bibinfo
  {pages} {013804} (\bibinfo {year} {2007})}\BibitemShut {NoStop}%
\bibitem [{\citenamefont {Larson}\ and\ \citenamefont
  {Irish}(2017)}]{Larson2017}%
  \BibitemOpen
  \bibfield  {author} {\bibinfo {author} {\bibfnamefont {Jonas}\ \bibnamefont
  {Larson}}\ and\ \bibinfo {author} {\bibfnamefont {Elinor~K}\ \bibnamefont
  {Irish}},\ }\bibfield  {title} {\enquote {\bibinfo {title} {Some remarks on
  ‘superradiant’ phase transitions in light-matter systems},}\ }\href
  {\doibase 10.1088/1751-8121/aa65dc} {\bibfield  {journal} {\bibinfo
  {journal} {J. Phys. A: Math. Theor.}\ }\textbf {\bibinfo {volume} {50}},\
  \bibinfo {pages} {174002} (\bibinfo {year} {2017})}\BibitemShut {NoStop}%
\bibitem [{\citenamefont {Guti\'errez-J\'auregui}\ and\ \citenamefont
  {Carmichael}(2018)}]{Gutierrez2018}%
  \BibitemOpen
  \bibfield  {author} {\bibinfo {author} {\bibfnamefont {R.}~\bibnamefont
  {Guti\'errez-J\'auregui}}\ and\ \bibinfo {author} {\bibfnamefont {H.~J.}\
  \bibnamefont {Carmichael}},\ }\bibfield  {title} {\enquote {\bibinfo {title}
  {{Dissipative quantum phase transitions of light in a generalized
  Jaynes-Cummings-Rabi model}},}\ }\href {\doibase 10.1103/PhysRevA.98.023804}
  {\bibfield  {journal} {\bibinfo  {journal} {Phys. Rev. A}\ }\textbf {\bibinfo
  {volume} {98}},\ \bibinfo {pages} {023804} (\bibinfo {year}
  {2018})}\BibitemShut {NoStop}%
\bibitem [{\citenamefont {Boneberg}\ \emph {et~al.}(2022)\citenamefont
  {Boneberg}, \citenamefont {Lesanovsky},\ and\ \citenamefont
  {Carollo}}]{Boneberg2022}%
  \BibitemOpen
  \bibfield  {author} {\bibinfo {author} {\bibfnamefont {Mario}\ \bibnamefont
  {Boneberg}}, \bibinfo {author} {\bibfnamefont {Igor}\ \bibnamefont
  {Lesanovsky}}, \ and\ \bibinfo {author} {\bibfnamefont {Federico}\
  \bibnamefont {Carollo}},\ }\bibfield  {title} {\enquote {\bibinfo {title}
  {Quantum fluctuations and correlations in open quantum {D}icke models},}\
  }\href {\doibase 10.1103/PhysRevA.106.012212} {\bibfield  {journal} {\bibinfo
   {journal} {Phys. Rev. A}\ }\textbf {\bibinfo {volume} {106}},\ \bibinfo
  {pages} {012212} (\bibinfo {year} {2022})}\BibitemShut {NoStop}%
\bibitem [{\citenamefont {Lyu}\ \emph {et~al.}(2024)\citenamefont {Lyu},
  \citenamefont {Kottmann}, \citenamefont {Plenio},\ and\ \citenamefont
  {Hwang}}]{Lyu2024}%
  \BibitemOpen
  \bibfield  {author} {\bibinfo {author} {\bibfnamefont {Guitao}\ \bibnamefont
  {Lyu}}, \bibinfo {author} {\bibfnamefont {Korbinian}\ \bibnamefont
  {Kottmann}}, \bibinfo {author} {\bibfnamefont {Martin~B.}\ \bibnamefont
  {Plenio}}, \ and\ \bibinfo {author} {\bibfnamefont {Myung-Joong}\
  \bibnamefont {Hwang}},\ }\bibfield  {title} {\enquote {\bibinfo {title}
  {Multicritical dissipative phase transitions in the anisotropic open quantum
  {R}abi model},}\ }\href {\doibase 10.1103/PhysRevResearch.6.033075}
  {\bibfield  {journal} {\bibinfo  {journal} {Phys. Rev. Res.}\ }\textbf
  {\bibinfo {volume} {6}},\ \bibinfo {pages} {033075} (\bibinfo {year}
  {2024})}\BibitemShut {NoStop}%
\bibitem [{\citenamefont {de~Aguiar}\ \emph {et~al.}(1992)\citenamefont
  {de~Aguiar}, \citenamefont {Furuya}, \citenamefont {Lewenkopf},\ and\
  \citenamefont {Nemes}}]{Deaguiar1992}%
  \BibitemOpen
  \bibfield  {author} {\bibinfo {author} {\bibfnamefont {M.A.M}\ \bibnamefont
  {de~Aguiar}}, \bibinfo {author} {\bibfnamefont {K}~\bibnamefont {Furuya}},
  \bibinfo {author} {\bibfnamefont {C.H}\ \bibnamefont {Lewenkopf}}, \ and\
  \bibinfo {author} {\bibfnamefont {M.C}\ \bibnamefont {Nemes}},\ }\bibfield
  {title} {\enquote {\bibinfo {title} {Chaos in a spin-boson system: Classical
  analysis},}\ }\href {\doibase https://doi.org/10.1016/0003-4916(92)90178-O}
  {\bibfield  {journal} {\bibinfo  {journal} {Ann. Phys.}\ }\textbf {\bibinfo
  {volume} {216}},\ \bibinfo {pages} {291 -- 312} (\bibinfo {year}
  {1992})}\BibitemShut {NoStop}%
\bibitem [{\citenamefont {Bastarrachea-Magnani}\ \emph
  {et~al.}(2014)\citenamefont {Bastarrachea-Magnani}, \citenamefont
  {Lerma-Hern\'andez},\ and\ \citenamefont {Hirsch}}]{Bastarrachea2014b}%
  \BibitemOpen
  \bibfield  {author} {\bibinfo {author} {\bibfnamefont {M.~A.}\ \bibnamefont
  {Bastarrachea-Magnani}}, \bibinfo {author} {\bibfnamefont {S.}~\bibnamefont
  {Lerma-Hern\'andez}}, \ and\ \bibinfo {author} {\bibfnamefont {J.~G.}\
  \bibnamefont {Hirsch}},\ }\bibfield  {title} {\enquote {\bibinfo {title}
  {Comparative quantum and semiclassical analysis of atom-field systems. {II}.
  {C}haos and regularity},}\ }\href {\doibase 10.1103/PhysRevA.89.032102}
  {\bibfield  {journal} {\bibinfo  {journal} {Phys. Rev. A}\ }\textbf {\bibinfo
  {volume} {89}},\ \bibinfo {pages} {032102} (\bibinfo {year}
  {2014})}\BibitemShut {NoStop}%
\bibitem [{\citenamefont {Ch\'avez-Carlos}\ \emph {et~al.}(2016)\citenamefont
  {Ch\'avez-Carlos}, \citenamefont {Bastarrachea-Magnani}, \citenamefont
  {Lerma-Hern\'andez},\ and\ \citenamefont {Hirsch}}]{Chavez2016}%
  \BibitemOpen
  \bibfield  {author} {\bibinfo {author} {\bibfnamefont {J.}~\bibnamefont
  {Ch\'avez-Carlos}}, \bibinfo {author} {\bibfnamefont {M.~A.}\ \bibnamefont
  {Bastarrachea-Magnani}}, \bibinfo {author} {\bibfnamefont {S.}~\bibnamefont
  {Lerma-Hern\'andez}}, \ and\ \bibinfo {author} {\bibfnamefont {J.~G.}\
  \bibnamefont {Hirsch}},\ }\bibfield  {title} {\enquote {\bibinfo {title}
  {Classical chaos in atom-field systems},}\ }\href {\doibase
  10.1103/PhysRevE.94.022209} {\bibfield  {journal} {\bibinfo  {journal} {Phys.
  Rev. E}\ }\textbf {\bibinfo {volume} {94}},\ \bibinfo {pages} {022209}
  (\bibinfo {year} {2016})}\BibitemShut {NoStop}%
\bibitem [{\citenamefont {Li}\ \emph {et~al.}(2022)\citenamefont {Li},
  \citenamefont {Fazio},\ and\ \citenamefont {Chesi}}]{Li2022}%
  \BibitemOpen
  \bibfield  {author} {\bibinfo {author} {\bibfnamefont {Jiahui}\ \bibnamefont
  {Li}}, \bibinfo {author} {\bibfnamefont {Rosario}\ \bibnamefont {Fazio}}, \
  and\ \bibinfo {author} {\bibfnamefont {Stefano}\ \bibnamefont {Chesi}},\
  }\bibfield  {title} {\enquote {\bibinfo {title} {Nonlinear dynamics of the
  dissipative anisotropic two-photon {D}icke model},}\ }\href {\doibase
  10.1088/1367-2630/ac8897} {\bibfield  {journal} {\bibinfo  {journal} {New J.
  Phys.}\ }\textbf {\bibinfo {volume} {24}},\ \bibinfo {pages} {083039}
  (\bibinfo {year} {2022})}\BibitemShut {NoStop}%
\bibitem [{\citenamefont {Li}\ and\ \citenamefont {Chesi}(2024)}]{Li2024}%
  \BibitemOpen
  \bibfield  {author} {\bibinfo {author} {\bibfnamefont {Jiahui}\ \bibnamefont
  {Li}}\ and\ \bibinfo {author} {\bibfnamefont {Stefano}\ \bibnamefont
  {Chesi}},\ }\bibfield  {title} {\enquote {\bibinfo {title} {Routes to chaos
  in the balanced two-photon {D}icke model with qubit dissipation},}\ }\href
  {\doibase 10.1103/PhysRevA.109.053702} {\bibfield  {journal} {\bibinfo
  {journal} {Phys. Rev. A}\ }\textbf {\bibinfo {volume} {109}},\ \bibinfo
  {pages} {053702} (\bibinfo {year} {2024})}\BibitemShut {NoStop}%
\bibitem [{\citenamefont {Vivek}\ \emph {et~al.}(2024)\citenamefont {Vivek},
  \citenamefont {Mondal}, \citenamefont {Chakraborty},\ and\ \citenamefont
  {Sinha}}]{Vivek2024ARXIV}%
  \BibitemOpen
  \bibfield  {author} {\bibinfo {author} {\bibfnamefont {G.}~\bibnamefont
  {Vivek}}, \bibinfo {author} {\bibfnamefont {Debabrata}\ \bibnamefont
  {Mondal}}, \bibinfo {author} {\bibfnamefont {Subhadeep}\ \bibnamefont
  {Chakraborty}}, \ and\ \bibinfo {author} {\bibfnamefont {S.}~\bibnamefont
  {Sinha}},\ }\href@noop {} {\enquote {\bibinfo {title} {Self-trapping
  phenomenon, multistability and chaos in open anisotropic {D}icke dimer},}\ }
  (\bibinfo {year} {2024}),\ \Eprint {http://arxiv.org/abs/2405.13809}
  {arXiv:2405.13809 [cond-mat.quant-gas]} \BibitemShut {NoStop}%
\bibitem [{\citenamefont {Markum}\ \emph {et~al.}(1999)\citenamefont {Markum},
  \citenamefont {Pullirsch},\ and\ \citenamefont {Wettig}}]{Markum1999}%
  \BibitemOpen
  \bibfield  {author} {\bibinfo {author} {\bibfnamefont {H.}~\bibnamefont
  {Markum}}, \bibinfo {author} {\bibfnamefont {R.}~\bibnamefont {Pullirsch}}, \
  and\ \bibinfo {author} {\bibfnamefont {T.}~\bibnamefont {Wettig}},\
  }\bibfield  {title} {\enquote {\bibinfo {title} {Non-{H}ermitian random
  matrix theory and lattice {QCD} with chemical potential},}\ }\href {\doibase
  10.1103/PhysRevLett.83.484} {\bibfield  {journal} {\bibinfo  {journal} {Phys.
  Rev. Lett.}\ }\textbf {\bibinfo {volume} {83}},\ \bibinfo {pages} {484--487}
  (\bibinfo {year} {1999})}\BibitemShut {NoStop}%
\bibitem [{\citenamefont {Ginibre}(1965)}]{Ginibre1965}%
  \BibitemOpen
  \bibfield  {author} {\bibinfo {author} {\bibfnamefont {Jean}\ \bibnamefont
  {Ginibre}},\ }\bibfield  {title} {\enquote {\bibinfo {title} {{Statistical
  ensembles of complex, quaternion, and real matrices}},}\ }\href {\doibase
  10.1063/1.1704292} {\bibfield  {journal} {\bibinfo  {journal} {J. Math.
  Phys.}\ }\textbf {\bibinfo {volume} {6}},\ \bibinfo {pages} {440--449}
  (\bibinfo {year} {1965})}\BibitemShut {NoStop}%
\bibitem [{\citenamefont {Anderson}\ and\ \citenamefont
  {Darling}(1952)}]{Anderson1952}%
  \BibitemOpen
  \bibfield  {author} {\bibinfo {author} {\bibfnamefont {T.~W.}\ \bibnamefont
  {Anderson}}\ and\ \bibinfo {author} {\bibfnamefont {D.~A.}\ \bibnamefont
  {Darling}},\ }\bibfield  {title} {\enquote {\bibinfo {title} {Asymptotic
  theory of certain ``goodness of fit'' criteria based on stochastic
  processes},}\ }\href {\doibase 10.1214/aoms/1177729437} {\bibfield  {journal}
  {\bibinfo  {journal} {Ann. Math. Stat.}\ }\textbf {\bibinfo {volume} {23}},\
  \bibinfo {pages} {193 -- 212} (\bibinfo {year} {1952})}\BibitemShut {NoStop}%
\bibitem [{foo()}]{footSM}%
  \BibitemOpen
  \href@noop {} {}\bibinfo {note} {See Supplemental Material
  at~\url{http://link.aps.org/supplemental/10.1103/PhysRevLett.133.240404},
  which includes Ref.~\cite{MukamelBook}, for additional information about the
  spectral analysis of the Dicke Liouvillian.}\BibitemShut {Stop}%
\bibitem [{\citenamefont {Oganesyan}\ and\ \citenamefont
  {Huse}(2007)}]{Oganesyan2007}%
  \BibitemOpen
  \bibfield  {author} {\bibinfo {author} {\bibfnamefont {Vadim}\ \bibnamefont
  {Oganesyan}}\ and\ \bibinfo {author} {\bibfnamefont {David~A.}\ \bibnamefont
  {Huse}},\ }\bibfield  {title} {\enquote {\bibinfo {title} {Localization of
  interacting fermions at high temperature},}\ }\href {\doibase
  10.1103/PhysRevB.75.155111} {\bibfield  {journal} {\bibinfo  {journal} {Phys.
  Rev. B}\ }\textbf {\bibinfo {volume} {75}},\ \bibinfo {pages} {155111}
  (\bibinfo {year} {2007})}\BibitemShut {NoStop}%
\bibitem [{\citenamefont {Atas}\ \emph {et~al.}(2013)\citenamefont {Atas},
  \citenamefont {Bogomolny}, \citenamefont {Giraud},\ and\ \citenamefont
  {Roux}}]{Atas2013}%
  \BibitemOpen
  \bibfield  {author} {\bibinfo {author} {\bibfnamefont {Y.~Y.}\ \bibnamefont
  {Atas}}, \bibinfo {author} {\bibfnamefont {E.}~\bibnamefont {Bogomolny}},
  \bibinfo {author} {\bibfnamefont {O.}~\bibnamefont {Giraud}}, \ and\ \bibinfo
  {author} {\bibfnamefont {G.}~\bibnamefont {Roux}},\ }\bibfield  {title}
  {\enquote {\bibinfo {title} {Distribution of the ratio of consecutive level
  spacings in random matrix ensembles},}\ }\href {\doibase
  10.1103/PhysRevLett.110.084101} {\bibfield  {journal} {\bibinfo  {journal}
  {Phys. Rev. Lett.}\ }\textbf {\bibinfo {volume} {110}},\ \bibinfo {pages}
  {084101} (\bibinfo {year} {2013})}\BibitemShut {NoStop}%
\bibitem [{\citenamefont {Tavis}\ and\ \citenamefont
  {Cummings}(1968)}]{Tavis1968}%
  \BibitemOpen
  \bibfield  {author} {\bibinfo {author} {\bibfnamefont {Michael}\ \bibnamefont
  {Tavis}}\ and\ \bibinfo {author} {\bibfnamefont {Frederick~W.}\ \bibnamefont
  {Cummings}},\ }\bibfield  {title} {\enquote {\bibinfo {title} {{Exact
  Solution for an $N$-Molecule---Radiation-Field {H}amiltonian}},}\ }\href
  {\doibase 10.1103/PhysRev.170.379} {\bibfield  {journal} {\bibinfo  {journal}
  {Phys. Rev.}\ }\textbf {\bibinfo {volume} {170}},\ \bibinfo {pages}
  {379--384} (\bibinfo {year} {1968})}\BibitemShut {NoStop}%
\bibitem [{\citenamefont {Garc\'{\i}a-Garc\'{\i}a}\ \emph
  {et~al.}(2024)\citenamefont {Garc\'{\i}a-Garc\'{\i}a}, \citenamefont
  {Verbaarschot},\ and\ \citenamefont {Zheng}}]{Garcia2024}%
  \BibitemOpen
  \bibfield  {author} {\bibinfo {author} {\bibfnamefont {Antonio~M.}\
  \bibnamefont {Garc\'{\i}a-Garc\'{\i}a}}, \bibinfo {author} {\bibfnamefont
  {Jacobus J.~M.}\ \bibnamefont {Verbaarschot}}, \ and\ \bibinfo {author}
  {\bibfnamefont {Jie-ping}\ \bibnamefont {Zheng}},\ }\bibfield  {title}
  {\enquote {\bibinfo {title} {Lyapunov exponent as a signature of dissipative
  many-body quantum chaos},}\ }\href {\doibase 10.1103/PhysRevD.110.086010}
  {\bibfield  {journal} {\bibinfo  {journal} {Phys. Rev. D}\ }\textbf {\bibinfo
  {volume} {110}},\ \bibinfo {pages} {086010} (\bibinfo {year}
  {2024})}\BibitemShut {NoStop}%
\bibitem [{\citenamefont {Mukamel}(1995)}]{MukamelBook}%
  \BibitemOpen
  \bibfield  {author} {\bibinfo {author} {\bibfnamefont {Shaul}\ \bibnamefont
  {Mukamel}},\ }\href@noop {} {\emph {\bibinfo {title} {Principles of Nonlinear
  Optical Spectroscopy}}}\ (\bibinfo  {publisher} {Oxford University Press},\
  \bibinfo {address} {New York, NY},\ \bibinfo {year} {1995})\BibitemShut
  {NoStop}%
\end{thebibliography}%

\newpage
\cleardoublepage

\begin{widetext}
    \begin{center}
    {\large \textbf{Supplemental Material: \\
Breakdown of the Quantum Distinction of Regular and Chaotic Classical Dynamics in Dissipative Systems \\}}
\end{center}

In this Supplemental Material, we extend the results shown in Figs.~1(c)-1(d) of the main text by providing the spectral analysis for different regions of the spectrum of the Dicke Liouvillian. In Sec.~I, we show examples of the level spacing distribution, $P(s)$, for the Liouvillians of the open anisotropic and isotropic Dicke models and give values of the Anderson-Darling statistical test, $A^2$, for various windows of the Liouvillian eigenvalues. In Sec.~II, we explain how the diagonalization of the Dicke Liouvillian is performed.

\section{SPECTRAL ANALYSIS OF THE DICKE LIOUVILLIAN}

\setcounter{equation}{0}

The Liouvillian $\hat{\mathcal{L}}$ of an open quantum system is, in general, a non-Hermitian operator with complex eigenvalues, 
\begin{equation}
    \hat{\mathcal{L}}|\lambda_{k}\rangle\rangle = \lambda_{k}|\lambda_{k}\rangle\rangle,
    \tag{\thesection.\theequation}\stepcounter{equation}
\end{equation}
where $\lambda_{k} \in \mathbb{C}$, as in the case of the Dicke Liouvillian, $\hat{\mathcal{L}}_{\text{D}}$. For these eigenvalues, the spacing can be defined as the minimum Euclidean distance in the complex plane between a reference eigenvalue $\lambda_{k}$ and its nearest neighbor $\lambda_{k}^{\text{NN}}$,
\begin{equation}
    s_{k}=|\lambda_{k}-\lambda_{k}^{\text{NN}}|.
    \tag{\thesection.\theequation}\stepcounter{equation}
\end{equation}

In the analysis of level statistics, the unfolding procedure to capture only local fluctuations is as important for complex spectra as for real spectra~\cite{HaakeBook,Markum1999,Akemann2019,Hamazaki2020}. In this work, we follow the unfolding procedure presented in Ref.~\cite{Akemann2019}.

To compare our numerical level spacing distributions $P(s)$ with analytical probability distributions, we use the Anderson-Darling statistical test~\cite{Anderson1952}. It is obtained by arranging the spacings in increasing order, $s_{k}\leq s_{k+1}$, and computing the parameter
\begin{align}
    \label{eqn:ADTest}
    A^{2} = - N - \sum_{k=1}^{N}\frac{2k-1}{N}\left(\ln[F(s_{k})] + \ln[1-F(s_{N+1-k})]\right),
    \tag{\thesection.\theequation}\stepcounter{equation}
\end{align}
where $F(s) = \int_{0}^{s}ds'P(s')$ is the cumulative distribution function of the probability distribution $P(s)$. If $A^{2}>2.5$, we can say with 95$\%$ of confidence that the data set does not come from the given probability distribution. In the present work, we perform the Anderson-Darling statistical test for the 2D Poisson and GinUE distributions presented in Eqs.~(10)-(11) of the main text.

\subsection{Numerical results}

The Fig.~\ref{fig:SM} below extends the analysis of level statistics presented in Figs.~1(c)-1(d) of the main text and reinforces that for both quantum open anisotropic and isotropic Dicke models, we find agreement with the GinUE distribution. The figure is obtained by diagonalizing the positive parity sector of the Dicke Liouvillian for the system parameters $\omega=\omega_{0}=\kappa=1$ and $j=1$ (see details in Sec.~II). For the truncation value $n_{\max}=70$, we have 11100 converged eigenvalues for the anisotropic case ($\gamma_{-} = 1$, $\gamma_{+} = 2$) and 9114 for the isotropic one ($\gamma_{-} = \gamma_{+} = \gamma = 2$). We organized the positive parity spectrum in increasing order of the absolute values of the eigenvalues, $|\lambda_{k}|\leq|\lambda_{k+1}|$, and the Anderson-Darling statistical test was performed by taking moving windows of consecutive eigenvalues.

Figure~\ref{fig:SM}(a) shows the Anderson-Darling statistical test for the anisotropic case ($\gamma_{-} = 1$, $\gamma_{+} = 2$) for moving windows of 500 eigenvalues covering the whole spectrum. The blue solid line represents the Anderson-Darling statistical test for the 2D Poisson distribution, while the red dotted line represents the same test for the GinUE distribution. The horizontal green dashed line marks the threshold of the Anderson-Darling statistical test, $A^{2}=2.5$. At low values of $|\lambda|$, there is no clear correspondence with a specific distribution, but for higher eigenvalues we verify agreement with the GinUE distribution. 

Figures~\ref{fig:SM}(a1)-\ref{fig:SM}(a2) show $P(s)$ for two specific cases corresponding to the region of eigenvalues indicated with the vertical black dashed lines in Fig.~\ref{fig:SM}(a), that is, we took the eigenvalues contained in the windows centered at $|\lambda|=26.4$ [Fig.~\ref{fig:SM}(a1)] and $|\lambda|=63.0$ [Fig.~\ref{fig:SM}(a2)]. For both cases, the spacing distribution (bars) coincides with the GinUE distribution (red dashed line), as ensured by the Anderson-Darling test values of $A_{\text{GinUE}}^{2}=0.88<2.5$ and $A_{\text{GinUE}}^{2}=0.73<2.5$, respectively.

\begin{figure}
    \centering
    \includegraphics[width=0.95\linewidth]{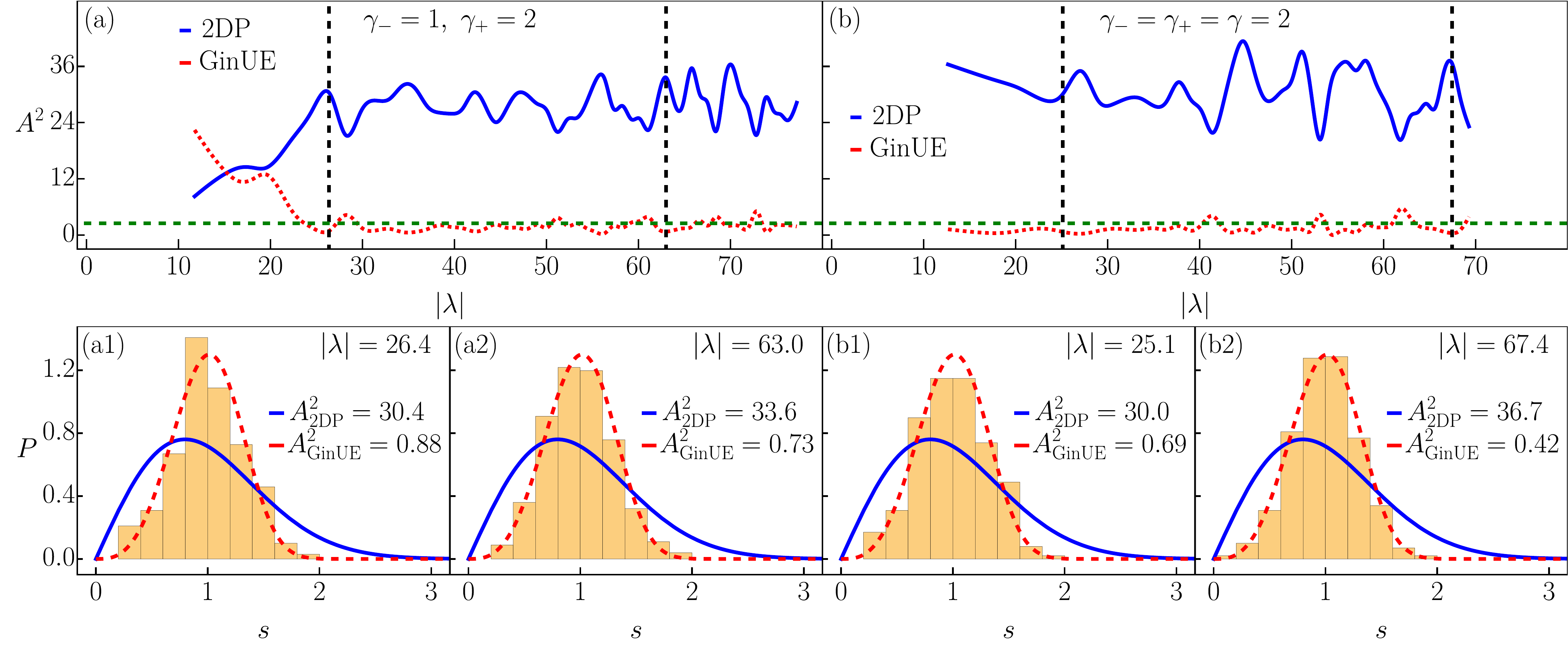}
    \caption{Anderson-Darling statistical test for the positive parity sector of the Dicke Liouvillian for (a) the anisotropic system with coupling strengths $\gamma_{-} = 1$ and $\gamma_{+} = 2$ and (b) the isotropic case with $\gamma_{-} = \gamma_{+} = \gamma = 2$. The blue solid (red dotted) line represents the Anderson-Darling parameter computed for the 2D Poisson (GinUE) distribution [see Eq.~\eqref{eqn:ADTest}]. The green dashed horizontal line marks the Anderson-Darling threshold, $A^{2}=2.5$. The black dashed vertical lines represent the mean value of two selected windows used in the spacing distributions, $P(s)$, shown in panels (a1)-(a2) and (b1)-(b2). The blue solid (red dashed) line in panels (a1)-(a2) and (b1)-(b2) represents the 2D Poisson (GinUE) distribution [see Eqs.~(10)-(11) of the main text]. System parameters: $\omega=\omega_{0}=\kappa=1$ and $j=1$.}
    \label{fig:SM}
\end{figure}

Figure~\ref{fig:SM}(b) shows the Anderson-Darling statistical test for the isotropic case ($\gamma_{-} = \gamma_{+} = \gamma = 2$), which indicates agreement with the GinUE distribution even for small values of $|\lambda|$. Figs.~\ref{fig:SM}(b1)-\ref{fig:SM}(b2) show the spacing distribution of two specific cases centered at $|\lambda|=25.1$ and $|\lambda|=67.4$, respectively. The agreement with the GinUE distribution is visible and ensured with the Anderson-Darling test values of $A_{\text{GinUE}}^{2}=0.69<2.5$ and $A_{\text{GinUE}}^{2}=0.42<2.5$, respectively. We notice that the fluctuations in the spectral tests can be reduced when selecting more eigenvalues or increasing the system size (see comparison for two system sizes, $j=1$ and $j=3$, in Ref. [63]).

\section{EXACT DIAGONALIZATION OF THE DICKE LIOUVILLIAN}

\setcounter{equation}{0}

To obtain a matrix representation of the Dicke Hamiltonian we use the Fock basis,
\begin{equation}
    |f\rangle = |n;j,m_{z}\rangle = |n\rangle\otimes|j,m_{z}\rangle ,
    \tag{\thesection.\theequation}\stepcounter{equation}
\end{equation}
which is a tensor product between the Fock states, $|n\rangle$, of the bosonic subspace and the angular momentum states, $|j,m_{z}\rangle$, of the atomic subspace. The global index $f=(2j+1)n+m_{z}+j+1$ orders the elements of the Hamiltonian matrix, where $n=0,\ldots,\infty$ and $m_{z}=-j,-j+1,\ldots,j-1,j$. The Hilbert space of the atomic subspace has dimension $2j+1$, while the Hilbert space of the bosonic subspace is infinite. The numerical diagonalization requires the truncation of the bosonic subspace to a chosen value of the number of photons, $n_{\max}$. Thus, the  truncated Hilbert space has dimension $\mathcal{D}_{\text{H}}=(2j+1)(n_{\max}+1)$.

To get a matrix representation of the Dicke Liouvillian, we use the tetradic notation~\cite{MukamelBook}. We use the Fock basis to generate the Liouville basis identified by the projectors
\begin{equation}
    |f',f\rangle\rangle = |f'\rangle\langle f| = |n';j,m'_{z}\rangle\langle n;j,m_{z}| ,
    \tag{\thesection.\theequation}\stepcounter{equation}
\end{equation}
where $n',n=0,\ldots,\infty$ and $m'_{z},m_{z}=-j,-j+1,\ldots,j-1,j$. In the equation above, the notation $|\bullet\rangle\rangle$ represents a state in the Liouville space. The truncated Liouville space for a chosen number of photons, $n_{\max}$, has dimension $\mathcal{D}_{\text{L}}=\mathcal{D}_{\text{H}}^{2}$.

The Liouvillian matrix is given explicitly by
\begin{align}
    L_{g'g,f'f}^{\text{D}} & = \langle\langle g',g|\hat{\mathcal{L}}_{\text{D}}|f',f\rangle\rangle = \text{Tr}\{|g\rangle\langle g'|\hat{\mathcal{L}}_{\text{D}}|f'\rangle\langle f|\} = \sum_{k}\langle k|g\rangle\langle g'|\hat{\mathcal{L}}_{\text{D}}|f'\rangle\langle f|k\rangle
    \tag{\thesection.\theequation}\stepcounter{equation} \\
    & = -i(\langle g'|\hat{H}_{\text{D}}|f'\rangle\delta_{f,g}-\langle f|\hat{H}_{\text{D}}|g\rangle\delta_{g',f'}) + \kappa(2\langle g'|\hat{a}|f'\rangle\langle f|\hat{a}^{\dagger}|g\rangle - \langle g'|\hat{a}^{\dagger}\hat{a}|f'\rangle\delta_{f,g} -\langle f|\hat{a}^{\dagger}\hat{a}|g\rangle\delta_{g',f'}) , \nonumber
\end{align}
where the matrix elements are
\begin{align}
    \label{eq:MatrixLiouvillian}
    \langle f'|\hat{H}_{\text{D}}|f\rangle = & \; (\omega n+\omega_{0}m_{z})\delta_{n',n}\delta_{m'_{z},m_{z}} + \frac{1}{\sqrt{\mathcal{N}}}\left(\gamma_{-}\sqrt{n}\delta_{n',n-1}+\gamma_{+}\sqrt{n+1}\delta_{n',n+1}\right)C_{m_{z}}^{+}\delta_{m'_{z},m_{z}+1} +
    \tag{\thesection.\theequation}\stepcounter{equation} \\
    & \; \frac{1}{\sqrt{\mathcal{N}}}\left(\gamma_{-}\sqrt{n+1}\delta_{n',n+1}+\gamma_{+}\sqrt{n}\delta_{n',n-1}\right)C_{m_{z}}^{-}\delta_{m'_{z},m_{z}-1} , \nonumber \\
    \langle f'|\hat{a}|f\rangle = & \; \sqrt{n}\delta_{n',n-1}\delta_{m'_{z},m_{z}} ,
    \tag{\thesection.\theequation}\stepcounter{equation} \\
    \langle f'|\hat{a}^{\dagger}|f\rangle = & \; \sqrt{n+1}\delta_{n',n+1}\delta_{m'_{z},m_{z}} ,
    \tag{\thesection.\theequation}\stepcounter{equation} \\
    \langle f'|\hat{a}^{\dagger}\hat{a}|f\rangle = & \; n\delta_{n',n}\delta_{m'_{z},m_{z}} ,
    \tag{\thesection.\theequation}\stepcounter{equation} \\
    \langle f'|\hat{a}\hat{a}^{\dagger}|f\rangle = & \; (n+1)\delta_{n',n}\delta_{m'_{z},m_{z}} ,
    \tag{\thesection.\theequation}\stepcounter{equation}
\end{align}
and $C_{m_{z}}^{\pm}=\sqrt{j(j+1)-m_{z}(m_{z}\pm1)}$. The isotropic Dicke Liouvillian is recovered from Eq.~\eqref{eq:MatrixLiouvillian} by using $\gamma_{-}=\gamma_{+}=\gamma$.

\subsection{Liouville basis with parity}

\setcounter{equation}{0}

The Fock basis is the eigenbasis of the parity operator $\hat{\Pi} = \text{exp}[i\pi(\hat{a}^{\dagger}\hat{a}+\hat{J}_{z}+j\hat{\mathbb{I}})]$, $\hat{\Pi}|f\rangle=p|f\rangle$, whose eigenvalues $p=(-1)^{n+m_{z}+j}=\pm1$ belong to one of the two sectors with well-defined parity for the Dicke Hamiltonian, $[\hat{H}_{\text{D}},\hat{\Pi}]=0$. Analogously, the Liouville basis is an eigenbasis of the parity superoperator $\hat{\mathcal{P}}$,
\begin{equation}
    \hat{\mathcal{P}}|f',f\rangle\rangle = \hat{\Pi}|f'\rangle\langle f|\hat{\Pi}^{\dagger} = P|f',f\rangle\rangle , \tag{\thesection.A.\theequation}\stepcounter{equation}
\end{equation}
where the eigenvalues $P=(-1)^{n'+m'_{z}-n-m_{z}}=\pm1$ belong to one of the two sectors of well-defined parity for the Dicke Liouvillian, $[\hat{\mathcal{L}}_{\text{D}},\hat{\mathcal{P}}]=0$.

\end{widetext}

\end{document}